\documentclass{ar-1col}
\renewcommand{\d}{\mathrm{d}}
\usepackage{amsmath,amssymb}
\usepackage[usenames,dvipsnames]{xcolor}

\newcommand{\LB}[1]{{\color{black} {#1}}}
\newcommand{\LBbis}[1]{{\color{black} {#1}}}
\newcommand{\NK}[1]{{\color{black} {#1}}}
\newcommand{\NKbis}[1]{{\color{black} {#1}}}

\newcommand{\R}[0]{{\cal{R}}}

\usepackage[comma]{natbib}
\usepackage{url}
\jyear{2021}

\begin{document}

\markboth{Kavokine, Netz and Bocquet}{Fluids at the Nanoscale}

\title{Fluids at the Nanoscale: from continuum to sub-continuum transport}

\author{Nikita Kavokine,$^1$ Roland R. Netz,$^2$ and Lyd\'eric Bocquet$^1$
\affil{$^1$Laboratoire de Physique de l'\'Ecole Normale Sup\'erieure, ENS, Universit\'e PSL, CNRS, Sorbonne Universit\'e, Universit\'e Paris-Diderot, Sorbonne Paris Cit\'e, Paris, France; email: nikita.kavokine@ens.fr, lyderic.bocquet@ens.fr}
\affil{$^2$Fachbereich Physik, Freie Universitat Berlin, Berlin 14195, Germany}
}

\begin{abstract}
Nanofluidics has firmly established itself as a new field in fluid mechanics, as novel properties have been shown to emerge in fluids at the nanometric scale. Thanks to recent developments in fabrication technology, artificial nanofluidic systems are now being designed at the scale of biological nanopores. This ultimate step in scale reduction has pushed the development of new experimental techniques and new theoretical tools, bridging fluid mechanics, statistical mechanics and condensed matter physics. This review is intended as a toolbox for fluids at the nanometre scale. After presenting the basic equations that govern fluid behaviour in the continuum limit, we will show how these equations break down and new properties emerge in molecular scale confinement.
\end{abstract}

\begin{keywords}
nanofluidics, slippage, ion transport, sub-continuum, non-mean-field, statistical mechanics
\end{keywords}
\maketitle

\section{INTRODUCTION}

Fluid flows at the nanometre scale have been studied indirectly in various disciplines for the last fifty years (\cite{Eijkel2005}). However, it is only fifteen years ago that nanofluidics has emerged as a field on its own, first as a natural extension of microfluidics towards smaller scales. Back then, it was an issue in itself to establish that nanofluidics deserves its own name, meaning that there are specific effects at the nano-scale that are not present at the micro-scale. 

Indeed, the "ultimate scale" for observing specific effects is set by the molecular size of the fluid ; more precisely, a critical confinement $\ell_c = 1~\rm nm $ has been generally accepted as the limit of validity for the Navier-Stokes equation (\cite{Bocquet2010,Sparreboom2010}). Moreover, it is at the molecular scale that the fluidic functions of biological systems emerge: from the giant permeability and perfect selectivity of the aquaporin (\cite{Murata2000}), to the ion specificity of KcsA channels (\cite{MacKinnon2004}), to the mechano-sensitivity of Piezo channels (\cite{Wu2017}), to name a few. However, ten years ago, 
the exploration of this ultimate scale was hindered by technical challenges, as molecular scale channels could not be fabricated artificially.

\LBbis{A decade later, }nanofluidics has firmly established itself as a field (\cite{Bocquet2020}). Indeed, lengthscales associated with the electrostatics and the fluctuations of surfaces may reach up to several tens of nanometres, and their effects may be probed specifically in systems without molecular scale confinement (\cite{Schoch2008,Bocquet2010,Sparreboom2010}). These lengthscales govern the key nanofluidic phenomena that have been demonstrated over the last ten years, such as, for instance, \LBbis{fast flows in carbon nanotubes (\cite{Holt2006})}, diffusio-osmotic energy conversion (\cite{Siria2013}) or diode-type effects~(\cite{Vlassiouk2007a}). 

However, the progress in fabrication technology has now allowed to overcome the challenges that have hindered the development of nanofluidics at the ultimate scales, and artificial devices with confinement down to about one water molecule size (3 $\overset{\circ}{\rm A}$) have been achieved, in 0D, 1D or 2D geometry (\cite{Feng2016b,Tunuguntla2017,Gopinadhan2019}). It is therefore an exciting time for nanofluidics, since it now has the potential to reverse-engineer biological functions: minimal artificial systems that mimic biological processes may be designed and studied. Furthermore, nanofluidics is known for its short path from fundamental science to applications and innovation, and developments in single channel fabrication are likely to have direct implications for filtration and membrane science (\cite{Bocquet2020}). 

\begin{figure}
\includegraphics[width=0.65\textwidth]{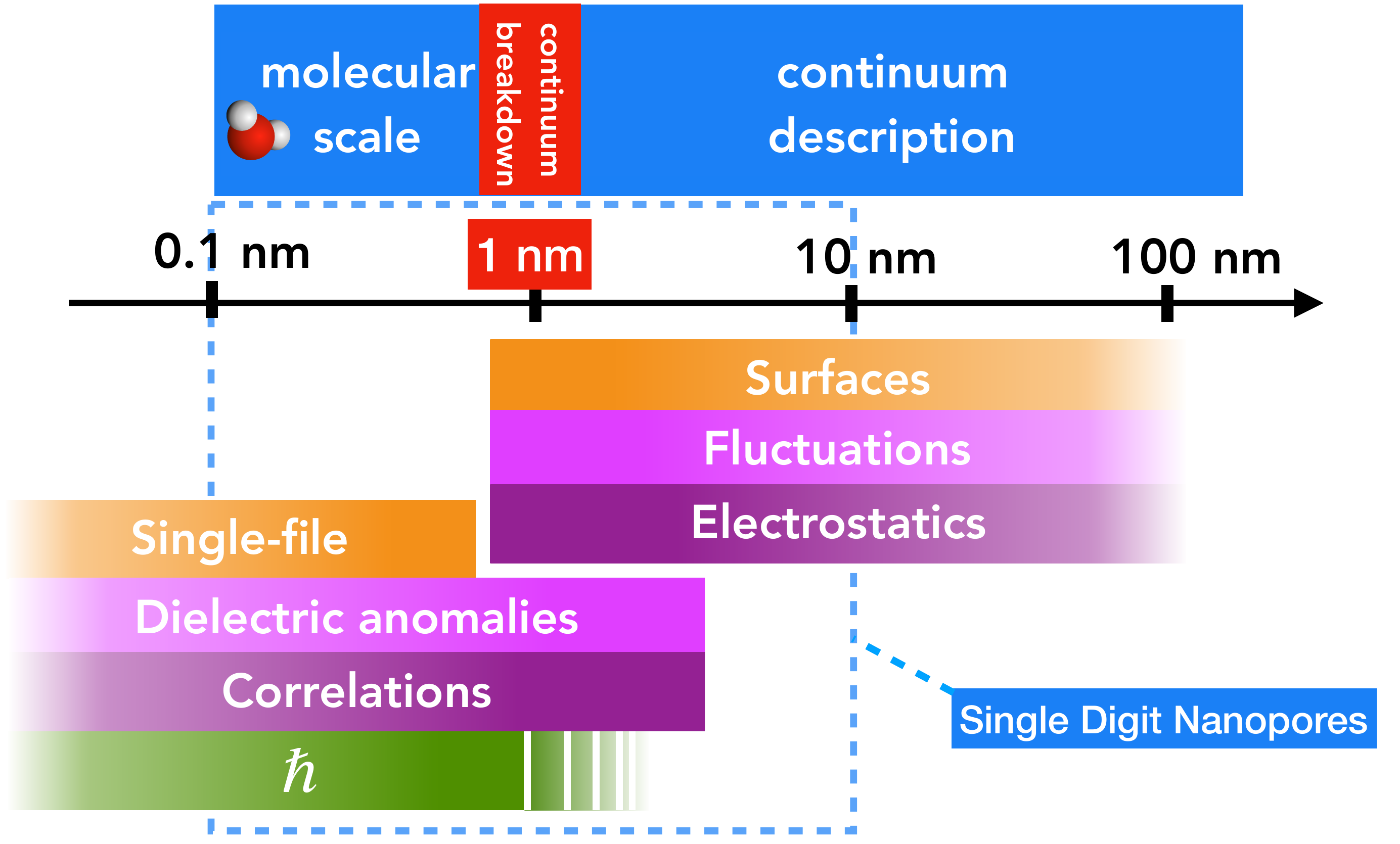}
\caption{An overview of nanofluidic lengthscales. Main ingredients of the physics above the continuum limit, and below.}
\end{figure}

At this time of accelerated development \LBbis{of the field}, one should realize that the \LBbis{nanoscales} under study require new tools for understanding the physics at play. The need for "new physics" is particularly emphasised in a recent review by \cite{Faucher2019}, which identifies "critical knowledge gaps in mass transport through single-digit nanopores" -- nanopores that are less than 10~nm in size. Those knowledge gaps \LBbis{exist} \NKbis{largely} because the tools applied for understanding the experiments \LBbis{are built on} macroscopic fluid mechanics and continuum electrostatics. But consider now a typical nanochannel of radius 1 nm. At physiological salt concentration, it contains only a single ion per 50 nm channel length. Similarly, the carbon nanotube porins studied by \cite{Tunuguntla2017} each contain about 30 water molecules. These numbers strongly suggest that discrete particle effects may be important, and call for a statistical mechanics description. Furthermore, below 1 nm confinement, the length scales associated with the fluid dynamics become comparable to characteristic length scales of the electrons in the confining solids, such as the Thomas-Fermi length (\cite{Mahan}). This points to the necessity of describing the confining solids at the level of condensed-matter physics, and not simply as a space impenetrable to the fluid molecules. Overall, understanding fluidic phenomena at the nanometer scales requires bridging the gap between fluid mechanics, statistical mechanics and condensed matter physics. 

In the past years, the breakdown of continuum equations has often set a hard limit for fluid mechanics: below the continuum limit was the realm of molecular simulations. 
However, the need for understanding experiments has pushed for the development (or rediscovery) of analytical tools that have allowed to identify some specific phenomena and associated length scales. The description of these phenomena is a key part of this review, as summarised in figure 1. 
 The review is organised as follows. In section 2, we give a brief overview of available nanofluidic systems and fabrication methods. In section 3, we focus on continuum modelling of nanofluidic systems, with particular emphasis on the precautions that should be taken when applying it to the smallest scales. Finally, in section 4, we go below the continuum limit, and highlight the specific phenomena that emerge along with the theoretical tools to describe them.

\section{THE TOOLBOX OF EXPERIMENTAL SYSTEMS}

Nanofluidics generally follows a bottom-up approach. Elementary phenomena are understood at the well-controlled scale of the individual channel, before eventually being applied to more complex systems. Hence, the design of these well-controlled systems is paramount to the development of the field. We start this review by going through the systems that have so far been achieved, in order of dimensionality. 

\begin{figure}
\includegraphics[width=\textwidth]{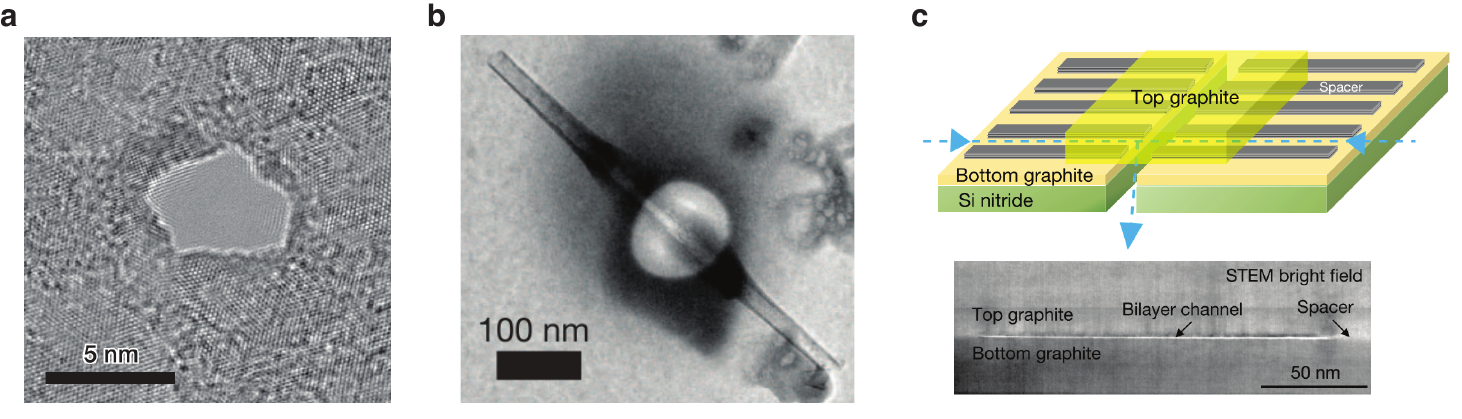}
\caption{State-of-the-art nanofluidic systems of various dimensions. \textbf{a.} TEM-drilled nanopore in single layer $\rm MoS_2$ (\cite{Feng2016b}). \textbf{b.} Boron nitride nanotube inserted into a SiN membrane (\cite{Siria2013}). \textbf{c.} Schematic and TEM image of the nanoslit device fabricated by \cite{Radha2016}.}
\end{figure}

\subsection{Nanopores}

Nanopores are channels whose length $L$ and diameter $d$ are both in the nanometre range (figure 2a). \LB{Initial studies focused on solid-state nanopores drilled through membranes made out of ceramics such as SiN of SiC (\cite{Keyser2006,Dekker2007}). More recently the advent of 2D materials, such as graphene, hexagonal boron nitride (hBN) or $\rm MoS_2$, allowed for the exploration of nanopores in atomically thin membranes (\cite{Garaj2010,Sahu2019})}. 
Essentially three types of fabrication pathways have been reported for well-controlled nanopores in 2D materials. 

\textbf{Drilling with an electron or a focused ion beam (FIB).} A single 5 nm pore drilled in monolayer graphene was first reported by \cite{Garaj2010}. \cite{Celebi2014} reported FIB drilling of arrays of nanopores in bilayer graphene, ranging from 7.6 nm to 1~$\mu\rm m$ in diameter. 

\textbf{Electrochemical etching. }\cite{Feng2015} reported the opening of pores in monolayer $\rm MoS_2$ when placed in a salt solution in between two electrodes. Applying a potential above the oxydation potential of $\rm MoS_2$ resulted in the gradual removal of single $\rm MoS_2$ units, thus creating an opening of controlled size. Nanopores down to 0.6~nm in diameter have been reported (\cite{Feng2016a}). 

\textbf{The use of intrinsic defects in 2D materials.} Large area membranes made of graphene or hBN  are known to exhibit defects in the form of pores, ranging in size from a few angstroms to 15~nm depending on conditions (\cite{Walker2017,OHern2012}), or such defects may be generated on purpose in smaller membranes using ultraviolet-induced oxidative etching (\cite{Koenig2012}). The chemical vapour deposition (CVD) graphene membranes produced by \cite{Jain2015} exhibited pores in the sub-2~nm range spaced by 70 to 100~nm. Placing the membrane on top of a 30-40~nm pore of a SiN membrane allowed to statistically isolate and study a single pore. 


\subsection{Nanotubes}

Nanotubes are cylindrical channels of diameter $d$ in the nanometre range, and length $L \gg d$ (figure 2b). They are typically made out of carbon, or the isoelectronic boron nitride. The nanotubes themselves, as a product of self-assembly, are readily available, but interfacing a nanotube to a fluidic system and avoiding leakage is still an experimental challenge. Three distinct strategies for addressing this challenge have been reported. 

\textbf{Building a microfluidic system on top of CVD-grown single-walled carbon nanotubes}. The systems may comprise one or several carbon nanotubes, typically of diameter 1 to 2 nm. The tubes generally have very high aspect ratio with lengths up to 0.5 mm (\cite{Lee2010,Choi2013,Yazda2017}), although similarly built systems with 20~$\mu \rm m$ long nanotubes have also been reported (\cite{Pang2011}). 

\textbf{Insertion of a multiwalled nanotube into a solid state membrane.} \cite{Siria2013} reported the fabrication of nanofluidic devices comprising a single boron nitride nanotube, inserted into a hole milled in a SiN membrane by direct nanomanipulation under SEM. The hole could be sealed \emph{in situ} by cracking of naphtalene induced by the electron beam. The method was later extended to carbon nanotubes (\cite{Secchi2016}), of 30 to 100 nm inner diameter and about 1~$\mu \rm m$ in length, and recently to smaller, 2 nm inner diameter, multiwall carbon nanotubes. 

\textbf{Insertion of nanotubes into a lipid membrane.} \cite{Liu2013} reported the insertion of very short (5 to 10 nm) and very narrow (0.8 to 2 nm in diameter) nanotubes into a supported lipid membrane. Nanotubes were brought in contact with the lipid bilayer thanks to a microinjection probe. Recently, \cite{Tunuguntla2017} reported the self-assembly of similar nanotubes, which they term carbon nanotube porins, into phospholipid vesicles. A single patch of membrane could also be isolated in order to study a single porin.

\subsection{Nanoslits down to angstr\"om confinements}

\LBbis{Slit-like channels with one dimension below tens of nanometers were first made using micro- and nano- fabrication techniques}. But recently,
\cite{Radha2016} reported the manufacturing of two-dimensional channels by van der Waals assembly of 2D materials (figure 2c). A few layers of graphene were used as spacers between two crystals of graphite, hBN or $\rm MoS_2$, allowing for atomically smooth channels of a few $\mu\rm m$ in length, 100 nm in width and down to 7 \AA~in height, that is the thickness of two graphene layers. Very recently (\cite{Gopinadhan2019}), water transport through one-graphene-layer thick (3.4 \AA) channels was reported.  

\vspace{.2cm}

This brief overview highlights that nanofluidics at the molecular scale is now a reality. Not only molecular scale confinement is possible, but the geometry of the confinement and the nature of the confining materials can also be tuned.

\section{NANOFLUIDICS IN THE CONTINUUM LIMIT}
\subsection{Liquid transport}
\subsubsection{Basic equations}

The two-centuries-old Navier-Stokes equation is remarkably robust at describing fluid flow down to the smallest scale, typically $\ell_c = 1~\rm nm$ for water in normal pressure and temperature conditions (\cite{Bocquet2010}). This length scale is essentially a lower bound for defining a fluid viscosity $\eta$. Indeed, in macroscopic fluid mechanics, the kinematic viscosity $\nu = \eta/\rho$, where $\rho$ is the mass density, plays the role of a diffusion coefficient for the fluid momentum. For such a diffusion coefficient to be defined, the time required for momentum to diffuse across the system, $\ell_c^2/\nu$, must be larger than the timescale of molecular motion, which is the microscopic origin of diffusion. A water molecule at a thermal agitation speed of $300~\rm m\cdot s^{-1}$ moves by its own size in $\tau_c = 10^{-12}~\rm s$, which defines a molecular time scale. Therefore, viscosity may be defined down to a system size
\begin{equation}
\ell_c \sim \sqrt{\nu \tau_c} \sim 1~\rm nm.
\end{equation}
Below this length scale, water structuring due to surfaces, memory effects and other sub-continuum phenomena come into play: these will be discussed in section 4. 
For water flow at $10~\rm nm$ length scales, the Reynolds number remains smaller than 0.1 up to fluid velocities of $10~\rm m\cdot s^{-1}$. Hence, in nanofluidic systems, inertial effects may be safely neglected, and the fluid flow is described by the Stokes equation: 
\begin{equation}
\eta \Delta v + f = \nabla p,
\end{equation}
where $p$ is the pressure and $f$ a body force, which may be due, for example, to the application of an electric field (see section 3.3).

\subsubsection{Boundary conditions} 

Stokes flow is often solved with no-slip boundary conditions: the velocity of the liquid is assumed to vanish at a solid-liquid interface. This is, however, a limiting case of the more general Navier partial slip boundary condition, which enforces that the viscous stress at the interface should be balancing the solid liquid friction force. Within linear response theory, the friction force is proportional to the liquid velocity. For a fluid flowing in the direction $x$ along a surface of normal $z$, the force balance per unit area writes
$\sigma_{xz} = \lambda v_x,$
with $[\sigma]$ the stress tensor and $\lambda$ the friction coefficient per unit area (expressed in $\rm N \cdot s \cdot m^{-3}$). For a Newtonian fluid, $\sigma_{xz} = \eta \partial_z v_x$, which allows to rewrite the Navier boundary condition as 
\begin{equation}
v_x = \left. b \frac{\partial v_x}{\partial z}\right|_{\rm wall},
\end{equation}
introducing the slip length $b = \eta/\lambda$. The slip length can be geometrically interpreted as the depth inside the solid where the linearly extrapolated fluid velocity profile vanishes. Accordingly, the no-slip boundary condition corresponds to $\lambda \to \infty$ or $b \to 0$. The effect of the partial slip condition is to simply shift the no-slip velocity profile by the slip velocity, which is not negligible roughly within a slip length from the wall. 
Since slip lengths \LBbis{up to} tens of nanometers have been measured on atomically flat \LBbis{(and hydrophobic)} surfaces, slippage is expected to play a crucial role in nanofluidics, and some of its effects will be discussed in the following sections. In the smallest channels, of size $R \ll b$, a perfect slip boundary condition may even be appropriate: the flow is then controlled by entrance effects. 

\subsubsection{Geometry and entrance effects}

Experimentally, the flow profile inside a nanofluidic channel can hardly be resolved, and one typically measures the total flow rate $Q$. Under a pressure drop $\Delta P$ and no-slip boundary conditions, the flow rate through a cylindrical channel of radius $R$ is given by the Hagen-Poiseuille formula: 
\begin{equation}
Q_c = \frac{\pi R^4}{8 \eta L} \Delta P. 
\end{equation}
This formula assumes a channel length $L \gg R$, and thereby neglects the effect of channel mouths on the flow rate. But the transition from a macroscopic reservoir to a nanoscale channel is a source of viscous dissipation, as the streamlines need to be bent in order for the fluid to enter the channel. These entrance effects may be examined by considering the flow through an infinitely thin nanopore, which is of interest in itself, given the geometry of certain nanofluidic devices (see section 2.1). This problem was addressed by \cite{Sampson1891}. For a nanopore of radius $R$ (and vanishing length) under pressure drop $\Delta P$, Sampson obtained the expression of the flow rate as
\begin{equation}
Q_p = \frac{R^3}{3 \eta} \Delta P.
\label{sampson}
\end{equation}
The scaling in Sampson's formula naturally emerges from a Stokes equation where the only lengthscale is $R$: $\eta \Delta v = \nabla p \Rightarrow \eta v/R^2 \sim \Delta P/R$, and the typical fluid velocity is $v \sim Q/R^2$. In order to estimate the flow rate through a channel taking into account entrance effects, one may simply add the hydrodynamic resistances of the pore ($\R_p$) and the channel ($\R_c$). \LBbis{If one writes $Q_c =(\Delta P)_c/ \R_c$ and $Q_p = (\Delta P)_p/\R_p$, then the entrance-corrected flow rate $Q_{pc}$ is obtained by imposing $Q_p=Q_c=Q_{pc}$ and $\Delta P=(\Delta P)_c+(\Delta P)_p$, so that}
\begin{equation}
Q_{pc} = \frac{\Delta P}{\R_h + \R_p} =  \frac{\pi R^4}{8 \eta L}\frac{ \Delta P}{1+ \frac{3\pi}{8}\frac{R}{L}}. 
\label{qpc}
\end{equation}
An exact computation (\cite{Dagan1982}) shows that the error made by this a priori crude approximation is less than 1\%. Equation~\ref{qpc} makes a continuous transition between the nanopore and nanochannel regimes, and shows that entrance effects are apparently negligible for channel lengths that exceed a few channel radii. 

\begin{figure}
\centering
\includegraphics[width=\textwidth]{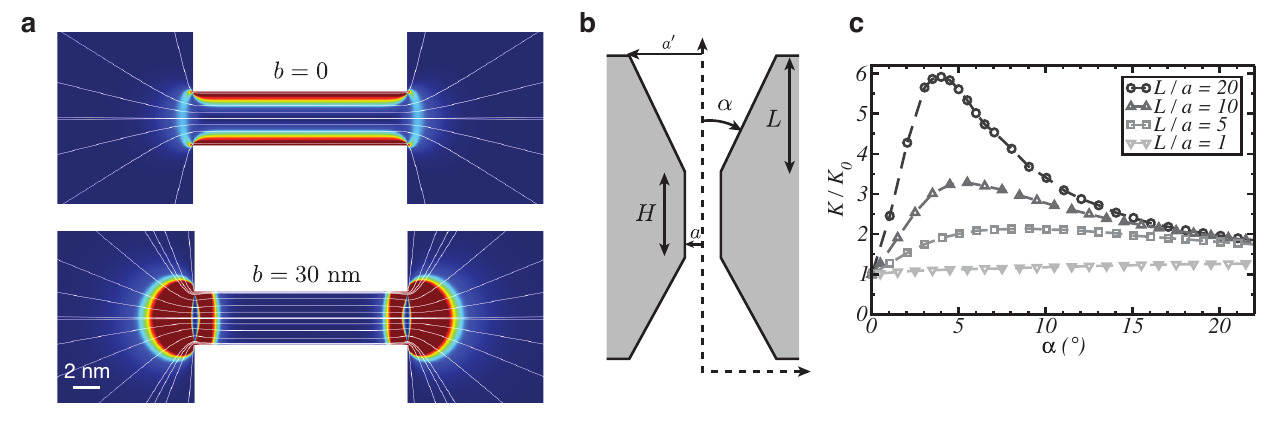}
\caption{Entrance effects in nanofluidics. \textbf{a.} Viscous dissipation rate, and streamlines, for the pressure-driven flow of water across a nanopore, as obtained from a finite elements solution of the Stokes equation (COMSOL). The colour scale, from blue to red, encodes the viscous dissipation. \textbf{b.} Geometric model of the aquaporin as considered by \cite{Gravelle2013}. \textbf{c.} Permeability of the model aquaporin as a function of the cone angle $\alpha$.}
\end{figure}

However, the above discussion has crucially not taken into account slippage, which, as we have highlighted in the previous section, is a strong effect at the nanoscale. Introducing a non-zero slip length $b$, the flow rate though a channel becomes 
\begin{equation}
Q_c = \frac{\pi R^4}{8 \eta L}\left( 1+ \frac{4b}{R}\right) \Delta P, 
\label{poiseuille}
\end{equation} 
while the flow rate through a pore is not significantly affected (\cite{Gravelle2013}), since the source of dissipation in that case is mostly geometric. \LBbis{A full expression can be obtained by gathering previous results, but in the limit where} $b \gg R$, the entrance-corrected flow rate becomes 
\begin{equation}
Q_{pc} = \frac{R^3}{3\eta} \frac{\Delta P}{1+ \frac{2L}{3\pi b}}.
\end{equation}
Thus, the hydrodynamic resistance is actually dominated by entrance effects as long as the channel is shorter than the slip length, rather than the channel radius (see figure 3a). 
In the presence of significant slippage, one should check whether the low Reynolds number assumption still holds. The average velocity through a channel of radius $R = 5~\rm nm$ and length $L = 1~\rm \mu m$, with slip length $b = 30~\rm nm$, under a pressure drop $\Delta P = 1~\rm bar$ is $v = 8~\rm mm \cdot s^{-1}$, which is 25 times faster than the no-slip result, but still well below the $1~\rm m\cdot s^{-1}$ threshold established in 3.1.1. 

Entrance-effect-dominated transport is particularly relevant for biological nanochannels, due to their relatively small aspect ratio. A striking example is aquaporin, which was recently studied from a hydrodynamic point of view by \cite{Gravelle2013}. Aquaporins are channel proteins that selectively transport water across the cell membrane. A simplified geometrical model for the aquaporin consists of two conical vestibules connected by a subnanometric channel where water flows in single file (figure 3b). The single-file transport is expected to be nearly frictionless, and therefore the limit to the aquaporin's permeability is set by the entrance effects in the conical vestibules. Moreover, since the channel in question is less than 1~nm wide and slip lengths of the order of 10~nm are expected, perfect slip boundary conditions are relevant for the flow, and the viscous dissipation has a purely geometric origin in the curvature of the stream lines. With the notations of figure 3b, Gravelle \emph{et al.} express the total hydrodynamic resistance of the aquaporin as 
\begin{equation}
R_{\rm AQP} = R_1 + R_2 = C_{\infty} \frac{\eta}{a'^3} + C_{\infty} \sin \alpha \frac{\eta}{a^3}. 
\end{equation}
Both terms are the analogue of Sampson's formula (eq.~\ref{sampson}). $C_{\infty} = 3.75$ replaces the factor 3 in the case of perfect slip boundary conditions, and the factor $\sin \alpha$ appears because at the cone-cylinder transition, the stream lines turn by an angle $\alpha$, as opposed to $\pi/2$ when the transition is from an infinite reservoir to a pore. With that, the permeability $K = R_{\rm AQP}^{-1}$ turns out to be a non-monotonous function on $\alpha$ (figure 3c), with the maximal permeability reached for cone angles in the range measured on aquaporin structures obtained by X-ray crystallography. It appears, therefore, that the geometry of the aquaporin is an optimum for hydrodynamic entrance effects, highlighting their particular relevance for nanoscale fluid transport.

\subsection{Gas transport}

\begin{figure}
\centering
\includegraphics[width=\textwidth]{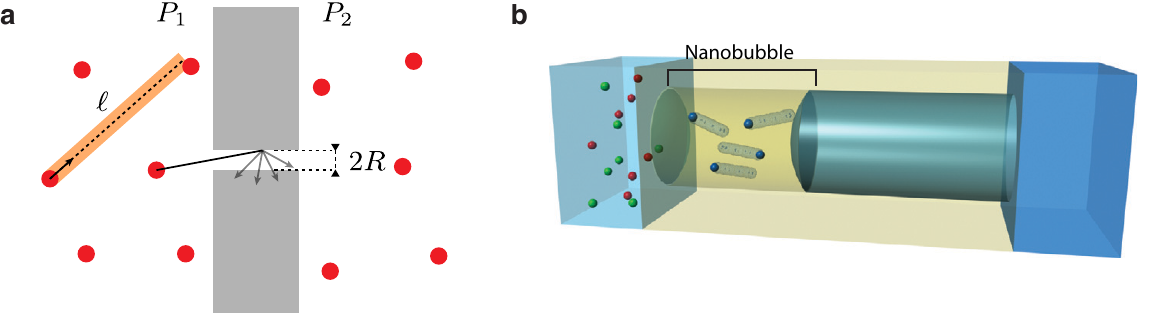}
\caption{\textbf{a.} Schematic of gas transport through a nanopore in the Knudsen regime $\ell \gg R$. A diffuse reflection at the pore wall is represented. \textbf{b.} Schematic of osmotic transport of water mediated by a nanobubble, reproduced from \cite{Lee2014}.}
\end{figure}

In a manner similar to liquid flow, gas flow can also be used to probe nanofluidic systems. Gas flow may display two limiting regimes, depending on the value of the Knudsen number, defined as 
\begin{equation}
Kn = \frac{\ell}{R},
\end{equation} 
where $\ell$ is the mean free path and $R$ is the typical system size. The mean-free path scales as the inverse density according to $\ell \sim (\rho \sigma^2)^{-1}$, with $\sigma$ the molecular diameter. For $Kn \ll 1$, the transport is dominated by intermolecular collisions and is therefore described by hydrodynamics; for example, by the Poiseuille or Sampson formulae introduced above. For $Kn \gg 1$, the transport is dominated by collisions with the walls and is described by molecular diffusion: this is the so-called Knudsen regime (\cite{Lei2016}). Consider a cylindrical channel of radius $R$ and length $L \gg R$, connecting two gas reservoirs at pressures $P_1$ and $P_2$ and same temperature $T$ (see figure 4a). 
One may then define a diffusion coefficient that relates the molecular flow rate $Q_K$ to the density gradient $\Delta n/L$ across the channel through a Fick-type law: 
\begin{equation}
\frac{Q_K}{\pi R^2} = D_K \frac{\Delta n}{L} = \frac{D_K}{k_BT} \frac{\Delta P}{L}.
\end{equation}
The last equality uses that $n = P/k_B T$ through the ideal gas law, and $\Delta P = P_1 - P_2$. The only parameters involved in the diffusion are the channel radius $R$ and the average thermal velocity $v^* = \sqrt{8k_B T/\pi m}$, where $m$ is the molecular mass of the gas. On dimensional grounds, the Knudsen diffusion coefficient should scale as $D_K \sim R v^*$. A kinetic theory computation (\cite{Knudsen1909,Steckelmacher1966}) yields $D_K = (2\pi/3)R v^*$, and the Knudsen formula:
\begin{equation}
Q_K = \frac{8}{3} \frac{\pi R^3}{\sqrt{2 \pi m k_B T}} \frac{\Delta P}{L}. 
\end{equation}
Now the Knudsen formula relies on the crucial approximation of diffuse reflexion on the channel walls: a molecule that hits a wall has a new velocity randomly picked out of a Boltzmann distribution. It was proposed by Maxwell that only a fraction $f$ of reflections should be diffuse, and the rest should be specular, that is correspond to elastic collisions. Smoluchowski then derived the corresponding correction to the Knudsen formula (\cite{Lei2016,V.Smoluchowski1910}): 
\begin{equation}
Q_K = \frac{2-f}{f} \cdot \frac{8}{3} \frac{\pi R^3}{\sqrt{2 \pi m k_B T}} \frac{\Delta P}{L}. 
\end{equation}
There is a divergence of the flow rate in the limit $f \to 0$, that is the channel opposes no more resistance to the gas flow when all the wall reflections are specular. This is to be expected, since in that case a molecule that enters the channel necessarily exits at the other end. The flow resistance is then dominated by entrance effects; in other words, the flow rate is given by the rate of molecules hitting the channel apertures. Gas molecules at a density $n$ hit an aperture of area $\pi R^2$ at a rate $(1/4)nv^* \pi R^2$. Hence the effusion flow rate through an opening of radius $R$ is
\begin{equation}
Q_e = \pi R^2 \Delta n \sqrt{\frac{k_B T}{2\pi m}} = \frac{\pi R^2}{\sqrt{2 \pi m k_B T}} \Delta P,
\end{equation}
which also sets the flow rate in a long channel with specular reflection at the walls. One notes there is a close analogy between the Poiseuille and Sampson formulae for liquids, on the one hand, and the Knudsen and effusion formulae for gases, on the other hand. The fraction of specular reflections $1-f$ plays a role similar to the slip length: if it is large, the transport in a long channel may still be dominated by entrance effects. 

The free effusion prediction has been verified in systems of graphene nanopores (\cite{Celebi2014}), with a transition to the hydrodynamic behaviour (Sampson formula) observed upon reducing the Knudsen number. Specular reflections have also been evidenced in longer channels. Gas flow exceeding the Knudsen prediction was measured in carbon nanotubes (\cite{Holt2006,Majumder2008}), and, recently, nearly ballistic transport was evidenced in angstrom-scale slits (\cite{Keerthi2018}). These results point out that the tendency to anomalously fast transport in nanoscale confinement exists not only for liquids, but also for gases. 

A striking example of gas-mediated osmotic flow was demonstrated by the group of Karnik. \cite{Lee2014} fabricated a nanoporous (70 nm pore size) membrane with partially hydrophobic pores, so that a nanobubble is trapped in each pore when the membrane is immersed in water (see figure 4b). The nanobubbles are impermeable to salt, but permeable to water through its vapour phase transport. They show that the membrane reaches an ion rejection of $99.9\%$, while competing with the permeability of state-of-the-art polyamide-based membranes. Such high permeability might seem counterintuitive since vapour phase transport is expected to scale with the density of water vapour, 1000 times lower than that of liquid water. However, using gas phase transport as the ion rejection mechanism allows for much larger pores (70 nm) then what would be required in the liquid phase ($\sim 1~\rm nm$) to ensure steric rejection of ions. The scaling of the water transport rate with the pore size cubed (\cite{Bocquet2014}) then explains the high transport efficiency in the gas phase. As such, nanoscale gas transport shows promise for addressing the permeability-selectivity tradeoff in membrane science.

\subsection{Ion transport}
The behaviour of ions in nanofluidic systems is of great practical interest with applications ranging from biological ion channels (\cite{MacKinnon2004}) to ionic liquids inside nanoporous electrodes (\cite{Chmiola2006,Merlet2012}). Ion transport also provides an indirect way of probing fluid transport, which is often useful, since electric currents are much easier to measure than fluid flow. However,  due to their long range Coulomb interactions and diffusive dynamics, ions in nanochannels give rise to physics of great complexity. 
In this section, we only sketch the main results with particular emphasis on how well-known equations apply to the smallest channels, and we refer refer the reader to other reviews on the subject (\cite{Bocquet2010,Schoch2008,Sparreboom2010}) for more details.

\subsubsection{Basic equations}
Consider an aqueous solution of monovalent salt. Let $\rho_+,\rho_-$ be the densities of positive and negative ions, respectively; $D$ the diffusion coefficient, here assumed to be the same for ions of either sign, and $\phi$ the electrostatic potential. In a mean-field treatment, the convective-diffusive dynamics of ions are described by a Smoluchowski equation: 
\begin{equation}
\frac{\partial \rho_{\pm}}{\partial t} = \nabla \cdot \left(D \nabla \rho_{\pm} \mp \frac{e D}{k_B T} (\nabla \phi) \rho_{\pm} + v \rho_{\pm}  \right),
\label{smoluchowski}
\end{equation}
where $e$ is the unit charge, $\phi$ the electrostatic potential and $v$ is the fluid velocity field. The mean-field assumption implies in particular that correlations between the ions can be neglected: the potential importance of such correlations in nanofluidics will be discussed in section 4. Until then, we proceed by specifying the electrostatic potential through Poisson's equation, 
\begin{equation}
\Delta \phi = -e \frac{\rho_+-\rho_-}{\epsilon}, 
\label{poisson}
\end{equation}
where $\epsilon$ is the dielectric permittivity of water. For now we assume it to be isotropic, though this assumption may break down for nano-confined water, as we discuss in section 4. Lastly, we specify the flow velocity through the Stokes equation, which now includes an electrostatic term: 
\begin{equation}
\eta \Delta v - e (\rho_+-\rho_-) \nabla \phi = \nabla p.
\label{stokes}
\end{equation}

We now apply these three coupled equations to a specific geometry, though the discussion that follows could be generalised to channels of any shpae. For simplicity, we consider a slit-like channel of height $h$, width $w$ and length $L$, with $w, L \gg h$, connecting two reservoirs of salt solution at concentration $\rho_s$, extending along the direction $x$, between $z = 0$ and $z = h$. When considering ion transport, it is important to note that most surfaces are charged in water, due either to the dissociation of surface groups or to the adsorption of ions (\cite{Perram1973,Grosjean2019,Mouhat2020}). We hence assume the channel wall carries a surface charge density $-\Sigma e$ ($\Sigma$ is expressed in elementary charges per unit area, and we assume here the surface charge to be negative). 

\subsubsection{Ionic conductance}

We first neglect the coupling of ion transport to water transport, and consider the electrophoretic (EP) contribution to the ionic current under an applied electric field $E$: this means that we start by setting the fluid velocity $v = 0$. In the steady state, the Smoluchowski equation (\ref{smoluchowski}) reduces to the so-called Nernst-Planck equations for the constant ionic fluxes (along the $x$ direction):
\begin{equation}
j_{\pm} = D \nabla_x \rho_{\pm} \mp \frac{e D}{k_B T} (\nabla_x \phi) \rho_{\pm}.
\label{nernstplanck}
\end{equation}
Together with the Poisson equation (\ref{poisson}), these constitute the widely used Poisson-Nernst-Planck (PNP) framework. In our geometry, the condition $L \gg h$ ensures that in the middle of the channel $\nabla_x \rho_{\pm} = 0$; moreover to first order in $E$, $\nabla_x \phi = -E$ and the densities reduce to their equilibrium values. Hence, the EP contribution to the ionic current writes 
\begin{equation}
I_{\rm ep} =  w\int_0^h \d z (j_+-j_-) = w\frac{e^2D}{k_B T} \int_0^h \d z (\rho_+ + \rho_-) E. 
\label{iep}
\end{equation}
In order to compute $I_{\rm ep}$, one needs to find the equilibrium solution of the coupled PNP equations for $\rho_+$ and $\rho_-$. 
At equilibrium $j_{\pm} = 0$ and the Nernst-Planck equations (\ref{nernstplanck}) can be integrated, imposing that in the reservoirs $\phi = 0$ and $\rho_{\pm} = \rho_s$. This yields a Boltzmann distribution for the ions in the electrostatic potential
\begin{equation}
\rho_{\pm} = \rho_s \exp\left(\mp \frac{e\phi}{k_BT} \right) \equiv \rho_s e^{\mp \psi},
\label{boltzmann}
\end{equation}
introducing a dimensionless potential $\psi$. Combining this with the Poisson equation (\ref{poisson}) yields the Poisson-Boltzmann (PB) equation: 
\begin{equation}
\Delta \psi - \lambda_D^{-2} \sinh(\psi) = 0, 
\label{PB}
\end{equation}
which introduces the Debye length $\lambda_D = (8 \pi \rho_s \ell_B)^{-1/2}$, with $\ell_B = e^2/(4\pi\epsilon k_B T)$ the Bjerrum length. 
In our geometry, the PB equation has an implicit solution in terms of an elliptic integral (\cite{Levine1975,Andelman1995}). We will not exploit it here, however, and we will instead recover the relevant limiting behaviours from qualitative considerations. 

It is well known that, roughly speaking, the Debye length sets the extension of the diffuse layer of counterions next to a charged surface (\cite{Israelachvili}). Hence, if the channel height $h \gg \lambda_D$, its two opposing walls do not 'see' each other. We expect the conductance to be the sum of a bulk term, and a surface term originating in the two Debye layers: 
\begin{equation}
I_{\rm ep} = 2 w \frac{e^2D}{k_B T} E(\rho_s h + \Sigma). 
\label{nooverlap}
\end{equation}
\LBbis{where $\rho_s h$ and $\Sigma$ account for the number of charge carriers in the bulk and at surfaces, respectively. }
In the opposite limit where $h \ll \lambda_D$, there is no more distinction between surface and bulk. All the quantities may be considered uniform across the channel: this is called the Debye overlap regime. However, one may not assume that the channel contains only counterions, and one should go back to the thermodynamic equilibrium with the reservoirs, which in this case bears the name of Donnan equilibrium (\cite{Bocquet2010}). One has $\rho_{\pm} = \rho_s e^{\mp \psi}$, which implies a chemical equilibrium $\rho_+ \rho_- = \rho_s^2$ in the channel. Going further, in the limit of long channel length, there should be local electroneutrality: $h(\rho_+-\rho_-) = 2\Sigma$. This yields
\begin{equation}
\rho_{\pm} = \sqrt{\rho_s^2 + (\Sigma/h)^2} \pm \Sigma/h, 
\label{rho_donnan}
\end{equation}
and the current-voltage relation in the Debye overlap is
\begin{equation}
I_{\rm ep} = 2 w \frac{e^2D}{k_B T} E\sqrt{(\rho_s h)^2 + \Sigma^2}. 
\label{overlap}
\end{equation}
Equation \ref{overlap} displays the first peculiarity of small channels: one may not simply add the surface and bulk contributions. Table 1 (see appendix) lists the values of Debye length for different electrolyte concentrations, showing that the Debye overlap regime is indeed relevant for experimentally accessible nanofluidic systems (see section 2). Qualitatively, eqs. (\ref{nooverlap}) and (\ref{overlap}) both predict saturation of the conductance at low salt concentrations at a value determined by the surface charge. The saturation occurs when $\rho_s \sim \Sigma/h$, which can be recast in the form $h \sim \rho_s/\Sigma \equiv \ell_{Du}$. $\ell_{Du}$ is called the Dukhin length and quantifies the competition between bulk and surface contributions to the conductance. For a channel narrower than $\ell_{Du}$, surface contributions dominate, and vice versa. The Dukhin length is going to be important in our upcoming discussion of entrance effects. 

\begin{marginnote}[]
\entry{Bjerrum length}{$\ell_B = \frac{e^2}{ 4\pi\epsilon_0k_BT}$. Distance between two unit charges at which their interaction energy is $k_BT$.}
\entry{Debye length}{$\lambda_D = (8 \pi \rho_s \ell_B)^{-1/2}$. Thickness of the diffuse layer of counterions next to a charged surface.}
\entry{Dukhin length}{$\ell_{Du} =  \Sigma/\rho_s$. Channel width below which surface conductance dominates over bulk conductance.}
\entry{Gouy-Chapman length}{$\ell_{GC} = (2\pi \Sigma \ell_B)^{-1}$. Distance a unit charge must travel from a charged surface so that its electrostatic energy is reduced by $k_BT$.}
\end{marginnote}

At this point, a remark should be made concerning the range of validity of equations (\ref{nooverlap}) and (\ref{overlap}). Indeed, they have been derived from qualitative considerations, without reference to the exact solution of the PB equation. Now, from eqs. (\ref{iep}) and (\ref{boltzmann}), one obtains more generally
\begin{equation}
I_{\rm ep} = 2 w \frac{e^2D}{k_B T} \left( \rho_s h \cosh(\psi(h/2)) + \frac{\mathcal{E}}{2k_BT} \right),
\label{iep_exact}
\end{equation}
where $\mathcal{E} = (\epsilon/2) \int_0^h (\partial_z \phi)^2 \d z$ is the electrostatic energy per unit area. The electric double layer can be pictured as a capacitor with charge $\Sigma$, hence one would expect its electrostatic energy to scale as $\Sigma^2$; \NKbis{moreover, this is the prediction of the linearised PB equation, i.e. eq. (\ref{PB}) with the approximation $\sinh \psi \approx \psi$.
This is in contrast to, eq. (\ref{nooverlap}), which predicts a linear scaling of the conductance with $\Sigma$: this scaling must therefore come from the non-linearities of the PB equation. 
The PB equation may be linearised if the potential varies by less than $k_B T$ across the Debye layer (or across the channel if there is Debye overlap). This is the case in the high concentration/low surface charge limit and specifically when the Debye length is smaller than the so-called Gouy-Chapman length: $\ell_{GC} = (2\pi \Sigma \ell_B)^{-1}$. On the other hand, Eq.~(\ref{overlap}), valid for the Debye overlap regime, is safe from a condition on $\ell_{GC}$, as it predicts both quadratic and linear scalings depending on the value of $\Sigma$, as long as there is Debye overlap.}

\subsubsection{Entrance effects}

Similarly to what we have discussed for liquid and gas transport, we may now ask, for ion transport, what is the additional electrical resistance due to the channel - reservoir interface. Equivalently, we may want to estimate the ionic conductance of a nanopore of small aspect ratio, say of radius $R$ and length $L\sim R$. The problem has first been considered in the context of biological channels by \cite{Hille1968} and \cite{Hall1975}. Hall solved the electrostatic problem with an electrode at infinity and an equipotential disk accounting for the entrance of the pore. Translating the solution into scaling arguments, the current through the pore entrance is $I \sim \pi R^2 \kappa_b \Delta V_{\rm out}/R$, where $\Delta V_{\rm out}$ is the voltage drop at the entrance of the pore, which is expected to occur over a distance $R$, and not over the macroscopic distance between the electrodes. \NK{This defines an electrical access resistance as the ratio $\Delta V_{\rm out}/I$}.
We now specialise to the thin Debye layer regime $\lambda_D \ll R$. If one simply sums the access resistance and the channel resistance as given by eq.~(\ref{nooverlap}) (adapted to cylindrical geometry), the current-voltage relation becomes
\begin{equation}
I_{\rm ep} = \kappa_b \left( \frac{L}{\pi R^2} \frac{1}{1+\ell_{Du}/R}+ \frac{1}{\alpha R} \right)^{-1} \Delta V, 
\end{equation}
with $\alpha$ a geometric factor which is 2 in Hall's computation, and introducing the bulk conductivity $\kappa_b  = 2e^2D\rho_s/(k_BT)$. The above equation predicts vanishing conductance as $\rho_s \to 0$, since the access resistance becomes infinite in this limit;  \LBbis{however, this is not what is observed experimentally in short nanopores (\cite{Lee2012,Feng2016b}).}
This inconsistency arises because, for a surface-charged pore, the access current has a surface contribution, in addition to the bulk contribution.
This surface contribution may be evaluated starting from charge conservation at the surface, which imposes a relation between the axial and radial components of the electric field, as pointed out by \cite{Khair2008}:
\begin{equation}
\kappa_b E_r = \partial_x [\kappa_s \theta(x) E_x],
\end{equation}
with $\kappa_s$ the surface conductance and $\theta$ the Heaviside function, accounting for the discontinuity of surface charge at the pore boundary, which leads to subtle consequences. \NK{Indeed, it reveals that the Dukhin length, $\ell_{Du} = \Sigma/\rho_s \sim \kappa_s/\kappa_b$}, is the relevant lengthscale for the surface contribution to the electric field outside the pore, instead of the channel radius \LBbis{or Debye length}. The Dukhin length appears here as an electrostatic healing length: 
 \LBbis{feeding the surface conduction at the nanopore mouth disturbs the electric field lines in the bulk over a length $\ell_{Du}$.} This interpretation is supported by the numerical results of \cite{Lee2012} as shown in figure 5a.
 \LBbis{The bending of the electric field lines can be estimated by a perturbative approach (\cite{Lee2012}) that leads to an analytical expression for the} corrected current-voltage relation: 
\begin{equation}
I_{\rm ep} = \kappa_b \left( \frac{L}{\pi R^2} \frac{1}{1+\ell_{Du}/R}+ \frac{1}{\alpha R+\beta \ell_{Du}} \right)^{-1} \Delta V.
\label{ion_entrance}
\end{equation}
The surface-charged pore therefore appears, from the perspective of entrance effects, as an uncharged pore of effective size $R + \ell_{Du}$, since the geometrical prefactor $\beta \approx 2$. In the limit of vanishing salt concentration, $\rho_s \to 0$, the conductance indeed saturates, and, as expected, the entrance correction disappears in the limit of large aspect ratio pores. 
Lee \emph{et al.} successfully compared the prediction of eq.~(\ref{ion_entrance}) to experimental measurements in SiN nanopores in the range $100 - 500~\rm nm$. More recently, it has been used to describe conductance measurements in $\rm MoS_2$ nanopores down to 2 nm in diameter (\cite{Feng2016b}). 

\begin{figure}
\centering
\includegraphics[width=0.8\textwidth]{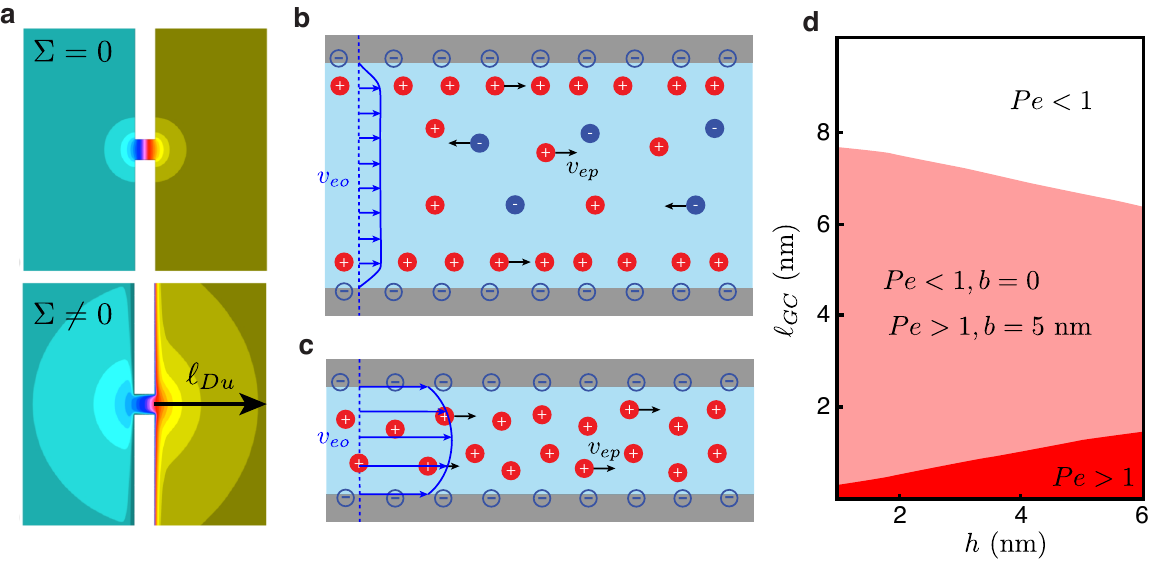}
\caption{\textbf{a.} Colour plot of the electrostatic potential around a nanopore immersed in a salt solution and subject to a voltage drop, in the charged and in the uncharged case. Around a charged pore, the lengthscale for variation of the potential is set by the Dukhin length. Adapted from \cite{Lee2012}. \textbf{b.} and \textbf{c.} Schematic representation of coupled ion and fluid transport in a nanochannel under electric field driving: in the thin Debye layer limit (\textbf{b.}) and in Debye overlap in the presence of slippage (\textbf{c.}). The electro-osmotic contribution dominates in the latter case. \textbf{d.} Peclet number (defined in the text), as a function of channel width and Gouy-Chapman length. The region in parameter space where $Pe >1$ is coloured in dark red in the no-slip case, and in light red when slippage is present.}
\end{figure}

\subsubsection{Coupling with fluid flow}
So far, we have neglected any coupling of ion transport to fluid transport. However, in the presence of charged surfaces, an external electric field exerts a net force on the charged Debye layer, which sets the fluid in motion. This interfacially-driven flow is termed electro-osmotic (EO) flow. The EO flow in turn drags along the ions in the Debye layer, which makes a supplementary contribution to the ionic current, that we denote $I_{\rm eo}$. This contribution has a convenient exact expression in the slit-like geometry considered in section 3.3.2. In analogy to eq.~(\ref{iep}), 
\begin{equation}
I_{\rm eo} = 2w\int_0^{h/2} e(\rho_+-\rho_-) v(z) \d z. 
\label{ieo}
\end{equation}
From the Poisson equation (\ref{poisson}), we may replace $e(\rho_+-\rho_-) = -\epsilon \partial^2_z \phi$. Then, integrating by parts, 
\begin{equation}
I_{\rm eo} = 2w\epsilon \left[ \int_0^{h/2} \partial_z \phi \partial_z v \, \d z + v(0) \partial_z \phi|_0 \right]. 
\end{equation}
Now, we may use the partial slip boundary condition $v(0) = b \partial_z v|_0$, as well as the electrostatic boundary condition $\partial_z \phi |_0 = e \Sigma/\epsilon$. This stems from the Gauss theorem applied to the surface, and the assumption that the medium outside the channel has much lower dielectric permittivity than water (\cite{Andelman1995}). Moreover, an integration of the Stokes equation~(\ref{stokes}) allows the replacement $ \partial_z v = -(\epsilon E/\eta) \partial_z \phi$. Altogether, we obtain
\begin{equation}
I_{\rm eo} = w\left[ \frac{2\epsilon}{\eta} \mathcal{E} + \frac{2 b e^2 \Sigma^2}{\eta} \right] E, 
\end{equation}
where $\mathcal{E} = (\epsilon/2) \int_0^h (\partial_z \phi)^2 \d z$ is the electrostatic energy per unit surface introduced in 3.3.1. It appears here that slippage has an additive contribution to the EO current, which strikingly does not depend on any electrolyte properties. 

We now consider the result in two limiting cases. First, in the thin Debye layer limit, by comparing with eq.~(\ref{iep_exact}), one may identify
\begin{equation}
I_{\rm eo} = \frac{k_B T}{2 \pi \ell_B \eta D} I_{\rm ep}^{\rm surf} + 2w\frac{b e^2 \Sigma^2}{\eta} E, 
\end{equation}
where $I_{\rm ep}^{\rm surf}$ is the surface contribution to the electrophoretic current. To quantify the importance of the EO contribution, one may compute the analogue of a Peclet number: 
\begin{equation}
Pe = \frac{I_{\rm eo}}{I_{\rm ep}^{\rm surf}} 
= \frac{3}{2} \frac{d_i}{\ell_B}\left(1+ \frac{b}{\ell_{GC}}\right). 
\end{equation}
We have introduced here the ion diameter $d_i$ by using the Einstein relation $D = k_B T/(3\pi \eta d_i)$ and we recall that $\ell_{GC} = (2\pi \Sigma \ell_B)^{-1}$. In water at room temperature, $d_i \sim \ell_B \sim 0.7~\rm nm$, therefore, in the absence of slippage, the EO contribution is of the same order as the surface EP contribution, $Pe \sim 1$. Table in the supplemental appendix lists, for reference, values of the Gouy-Chapman length for typical surface charge values. As these are generally in the nanometre range, even in the case of moderate slippage ($b \sim 10~\rm nm$), there is a strong enhancement of the EO contribution. The threshold confinement below which the resulting surface contribution dominates over the bulk contribution is given by a rescaled Dukhin length: $\ell_{Du}^* = (b/\ell_{GC}) \ell_{Du}$. 

Second, in the Debye overlap regime, the EO current is readily determined from eq.~(\ref{ieo}), since the ion densities may then be considered uniform across the channel and are given by eq.~(\ref{rho_donnan}). Uniform ionic densities also imply that the flow has no longer a surface, but rather a volume driving. It is simply a Poiseuille flow, with the pressure gradient $\Delta P/L$ replaced by the electric driving force $e(\rho_+-\rho_-)E$. Altogether one obtains 
\begin{equation}
I_{\rm eo} \simeq \frac{w h}{3 \eta} e^2 \Sigma^2 \left(1+\frac{6b}{h}\right) E.
\end{equation}
Since in Debye overlap it makes no more sense to distinguish a surface and a bulk contribution, we define the Peclet number as the ratio of the EO current to the total EP current: 
\begin{equation}
Pe = \frac{I_{\rm eo}}{I_{\rm ep}} = \frac{d_i}{\ell_B} \frac{h}{4 \ell_{GC}} \frac{1+6b/h}{\sqrt{1+(h/\ell_{Du})^2}}. 
\end{equation}
Figure 5d illustrates the dependence of this Peclet number on $h$ and $\ell_{GC}$, with $\rho_s = 10^{-2}~\rm M$ so that there is Debye overlap. 
In the absence of slippage, the EO contribution dominates only for high surface charges. However, if a  small slip length $b = 5~\rm nm$ is introduced, the Peclet number exceeds 1 for all reasonable surface charges. Indeed, the Peclet number is then essentially determined by the ratio $b/\ell_{GC}$. This highlights that in the Debye overlap regime, ionic conduction should be mostly driven by electro-osmosis. 

If strong EO flows are expected, then the reciprocal effect, streaming current, is expected to be large as well. The streaming current results from the application of a pressure gradient $\Delta P/L$: in Debye overlap, the charge density $2\Sigma/h$ is simply dragged along by the Poiseuille flow. The current reads
\begin{equation}
I_{\rm str} \simeq hw \frac{eh\Sigma}{6\eta} \left(1+\frac{6b}{h}\right) \frac{\Delta P}{L} \equiv hw \frac{-\epsilon \zeta}{\eta} \frac{\Delta P}{L},
\end{equation}
where we have phenomenologically defined the zeta potential ($\zeta$) via the streaming mobility. Independent measurements of the surface charge through the voltage-driven current and of the zeta potential from the pressure-driven current may allow in principle to estimate the slip length $b$. Such an estimate does not replace a direct measurement, however, in particular because the slip length may directly depend on surface charge (\cite{Joly2020});  and \LBbis{additional charge-surface coupling effects may also occur}, as we discuss below. 

\subsubsection{Surface modifications to PNP theory}
 
A remark is in order at this point concerning the nature of surface charge, that we have so far assumed to be constant, whatever the conditions. Usually, surface charge is considered to result from the acid base reactivity of the surface when dipped into water, of the type $\rm [AH]_s \to [A^-]_s + H^+$, where the negative group $\rm [A^-]_s$ remains fixed on the surface, while the proton diffuses in solution. \NK{In the case of the air-water interface, surface charge may result from the adsorption of hydronium ions (\cite{Mamatkulov2017}), or charged impurities. The adsorption of surfactant impurities was also proposed as a charging mechanism for general hydrophobic surfaces (\cite{Uematsu2019}). Graphite and hBN, which are of particular relevance in nanofluidics, have {\it a priori} no obvious acid-base reactivity in water, but they may  develop a surface charge, through chemisorption or physisorption of hydroxyde ions, as inferred from experiments (\cite{Secchi2016a,Siria2013}), and recently confirmed by \emph{ab initio} simulations (\cite{Grosjean2019}).} Such a picture of surface charge implies that it may actually depend on electrolyte concentration, which is an example of {charge regulation}. Qualitatively, the salt concentration affects the surface potential, and therefore the concentration of $\rm H^+$ and $\rm OH^-$ ions at the surface, which in turn affects the chemical (or the adsorption) equilibrium which governs the surface charge. Such charge regulation has been invoked to explain several experiments on carbon nanotubes, where a scaling $I \propto \rho_s^{1/2}$ or $I \propto \rho_s^{1/3}$ was observed (\cite{Secchi2016a,Liu2010,Pang2011}). Various models have been developed (\cite{Secchi2016a,biesheuvel2016,manghi2018,Uematsu2018}), predicting \LBbis{a rich panorama of} different sublinear scalings of conductance with salt concentration, depending on the conditions. 

A further step that may be required to accurately describe the surface charge is to take into account its mobility. Surfaces charges may be mobile when embedded in lipid bilayers,
or with more relevance to nanofluidics, when resulting from adsorbed ions. For instance, the simulations of Grosjean et al. showed that physisorbed hydroxyde ions on graphene surfaces retain a high lateral mobility (\cite{Grosjean2019}). The effect of a mobile surface charge on ion and fluid transport coefficients (in the thin Debye layer regime) was \NK{the subject of several investigations (\cite{Maduar2015,Silkina2019,Mouterde2018})}. \LB{Using the framework of \cite{Mouterde2018}, one may} introduce two new friction coefficients: $\lambda_-$ between the adsorbed (negative) ions and the wall, and $\xi_-$ between the adsorbed ions and the fluid, in addition to the water-wall friction coefficient $\lambda$. The force balance on the interfacial fluid layer results in a modified partial slip boundary condition \LBbis{involving the tangential electric field at the surface:}
\begin{equation}
b_{\rm eff} \partial_z v\vert_{z=0} = v\vert_{z=0} -e \frac{\xi_-}{\xi_-+\lambda_-} \frac{b_{\rm eff}}{\eta} \Sigma (- \partial_x \phi)\vert_{z=0},
\end{equation}
where the effective slip length is 
\begin{equation}
b_{\rm eff} = \frac{b}{1+\frac{\lambda_- \xi_-}{\lambda(\lambda_-+\xi_-)}\Sigma}.
\end{equation}
The slip length is therefore reduced by the surface charge mobility, and the boundary condition involves an extra electric term. This has in general a moderating effect on the transport coefficients. For example, in the case of streaming current, the fluid brings along surfaces charges in addition to the counterions, so that the total current is reduced. Notable exceptions are diffusio-osmotic mobility and conductance: the overall ionic conductance increases when surface charges are able to move in response to the electric field. 

We have just introduced ion-wall and ion-water friction for adsorbed ions in the interfacial layer. Now, for the smallest accessible channels, which are comparable to the ion size (\cite{Radha2016,Lee2010}), such friction needs to be taken into account for all the ions, and should appear at the level of the transport equations. This was proposed, for instance, to rationalise the non-linear voltage-pressure couplings observed in angstrom-scale slits by \cite{Mouterde2019}. They observed that the streaming mobility $\mu$, defined by $I_{\rm str} = hw \mu \Delta P/L$, depends on the applied voltage in a qualitatively different way for graphite and BN slits. In the proposed 'surface PNP' model, the ion-wall ($\lambda_{\pm}$) and ion-water ($\xi_{\pm}$) friction coefficients are introduced in the same way as above, except they no longer apply to adsorbed 'surface charge' ions, but to all the regular salt ions. The equivalent of the Smoluchowski and Stokes equations need then to be rederived by considering the force balance on a single ion and on an element of fluid, respectively, as shown in the Appendix.

There is a rich phenomenology to be harnessed from the new couplings that appear when all components of the nanofluidic system interact with the surfaces. 
Nevertheless, such surface PNP theory is still in its infancy, and it requires input from more microscopic modelling in order to estimate the various friction coefficients it introduces. It even raises mathematical difficulties in the writing of a one-dimensional Poisson equation, whose Green's function has infinite range. Finally, the relevance of this phenomenological theory should be assessed in light of the correlation and structuring effects -- specific to the smallest scales -- that will be described in the next section. 

\section{BEYOND THE CONTINUUM DESCRIPTION}

In this section, we focus on the specific nanoscale effects that are not described in the framework of continuum hydro- and electro-dynamics. These effects mainly originate from thermal fluctuations, interparticle correlations and structuring effects, and, as such, they require theoretical tools that bridge the gap between physics of continuous media and statistical mechanics.

\subsection{Fluctuations}

\subsubsection{Fluctuating hydrodynamics}

The relative particle number fluctuations in an open system of $N$ particles scale as $1/\sqrt{N}$. In a nanopore of radius $1~\rm nm$ and length $10~\rm nm$, there are on average about 1000 water molecules and these fluctuations \LBbis{are of a few percent and} already not negligible. Fundamentally, the description of such fluctuations requires a statistical mechanics framework, such as density functional theory (\cite{Barrat_Hansen}).
\LB{However, these} approaches remain quite formal \LBbis{and difficult to implement,} particularly for \LB{finite length pores which are} non translationally-invariant systems. 

A simpler coarse-grained approach to fluctuations -- the so-called fluctuating hydrodynamics (FH) -- has been introduced by \cite{Landau_Lifshitz}. In fluctuating hydrodynamics, a random stress tensor $S$ is added to the Navier-Stokes equation:
\begin{equation}
\rho \partial_t \mathbf{v} + \rho (\mathbf{v} \cdot \nabla) \mathbf{v} = - \nabla p + \eta \Delta \mathbf{v} + \nabla \cdot S,
\end{equation}
where $S$ satisfies a fluctuation-dissipation theorem: $\langle S_{ij} (\mathbf{r},t) S_{kl}(\mathbf{r}',t') \rangle= 2 \eta k_B T (\delta_{ij} \delta_{kl} + \delta_{il} \delta_{jk})\delta(\mathbf{r}-\mathbf{r}')\delta(t-t')$, similarly to the fluctuating force in a Langevin equation; \LB{$\rho$ is the mass density of the fluid}. This description introduces thermal fluctuations in a setting where, otherwise, a continuum hydrodynamic description holds. FH has been extensively discussed in the literature (\cite{Hauge1973,Fox1970,Mashiyama1978}) and we only consider here its basic implications for fluid transport in a nanopore. 

In standard hydrodynamics, the fluid inside a nanopore is at rest when subject to no external force. The effect of fluctuations is to induce a stochastic flow through the nanopore, which can be pictured as a stochastic motion of the fluid centre of mass \LBbis{(CM)}. \cite{Detcheverry2012} solved the FH equations for the velocity correlation function, and found that the CM motion was described by a non-Markovian Langevin equation of the type 
\begin{equation}
m \frac{\d v}{\d t} = - \int_{-\infty}^t \d t'\xi(t-t') v(t') + F(t),
\label{langevin}
\end{equation}
where the random force $F(t)$ and the memory kernel $\xi$ are related by $\langle F(t) F(0) \rangle = 2 k_B T \xi(t)$. The memory kernel is indeed non-trivial (i.e. not reduced to a $\delta$ function), as confirmed by molecular dynamics (MD) simulations (\cite{Detcheverry2013}). Memory effects in the diffusion of fluid inside a nanopore are indeed expected from the analysis of relevant timescales. A nanopore of radius $R$ and length $L$ contains a mass $m = \rho \pi R^2 L$ of fluid, and its friction coefficient on the wall \LB{can be calculated from the Navier-Stokes equation as} $\xi_s \approx 8 \pi \eta L$. The velocity of the fluid CM then relaxes on a timescale $\tau \sim m/\xi_s \sim R^2/\nu$. But $R^2/\nu$ is the time required for fluid momentum to diffuse across the nanopore, that is the time for the wall friction force to establish itself. Therefore, the friction force may not adapt instantaneously to the CM velocity, and memory effects are to be expected. 

\NK{Properties that are determined by short timescale dynamics may be affected by those memory effects as it was shown to be the case for solute mobility in confinement (\cite{Daldrop2017}). More generally, memory effects play an important role in barrier-crossing processes (\cite{Kappler2018}).
But other properties only depend on the long timescale dynamics. }In particular, the fluid centre of mass diffusion coefficient, given by $\mathcal{D} = k_B T/(2 \int_0^{\infty} \xi(t) \d t)$, reduces to the Einstein expression $\mathcal{D} = k_B T/ \xi_s$, that would be expected if the diffusion was Markovian. Similarly, the hydrodynamic slip length $b$ is found to satisfy a Green-Kubo relation whatever the form of the memory kernel $\xi(t)$ (\cite{Bocquet2013}). 

From a practical perspective, one may evaluate the diffusion coefficient $\cal D$. From the Einstein relation, and the assumption of Poiseuille flow with slip length $b$, one obtains
\begin{equation}
\mathcal{D} = \frac{k_B T}{\xi} = \frac{k_B T}{\frac{8 \pi \eta L}{1+4b/R}+3\pi^2 \eta R},
\label{Dcm}
\end{equation}
\NK{where, in the most general case, the friction coefficient $\xi$ has two contributions, one from the channel interior and one from the entrance effects (\cite{Detcheverry2012}).}
Overall, the fluid CM diffusion $\mathcal{D}$ 
can be seen as a supplementary contribution to the diffusion coefficient of a particle inside the nanopore. If the size of the particle is comparable to that of the nanopore, its self-diffusion is strongly hindered (\cite{RENKIN1954}) and the fluid contribution may actually dominate the particle diffusion (\cite{Detcheverry2012}). 

\subsubsection{Noise}

Thermal fluctuations are at the origin of the noise in ionic current measurements through a nanopore. At zero applied voltage, the total noise amplitude satisfies the Nyquist relation $\langle I^2 \rangle = 2k_B T G$, where $G$ is the nanopore conductance. The electro-osmotic term in the conductance makes a supplementary contribution to the noise, which can actually be traced back to the hydrodynamic fluctuations of the fluid CM described in the previous paragraph (\cite{Detcheverry2012}). 

When resolved in frequency, numerous experiments on artificial nanochannels (\cite{Smeets2008,Siwy2002,Hoogerheide2009,Tasserit2010,Secchi2016a}) and biological pores (\cite{Wohnsland1997,Bezrukov2000}), have shown that the current spectrum exhibits 'pink noise', that is noise that scales with frequency $f$ as $f^{-\alpha}$, with $\alpha$ in the range $0.5 - 1.5$. Such a spectrum is traditionally described by the empirical Hooge's law (\cite{Hooge1969}): 
\begin{equation}
\langle \delta I^2 \rangle (f) = A \frac{\langle I^2 \rangle}{f^{\alpha}},
\end{equation}
with $A$ a coefficient inversely proportional to the number of charge carriers. Notably, in experiments, the $f^{-\alpha}$ scaling is observed down to frequencies below 1 Hz, showing that some correlations in the system under study exist even at such low frequencies. Correlations between ions have been put forward as a possible origin for the pink noise (\cite{Zorkot2016}). On the other hand, \cite{Gravelle2019} demonstrated in MD simulations that pink noise persisted even with non-interacting ions, if reversible ion adsorption was allowed on the channel wall. For simple diffusive dynamics, the longest correlation time that may be expected for adsorption-desorption processes is $R^2/D$, with $R$ the pore radius and $D$ the ion diffusion coefficient. For a 10 nm pore, this would correspond to a frequency \LB{cutoff} of 100 MHz, below which the noise spectrum should be flat.
\LB{Such a high cutoff frequency} is in clear contradiction with experiment. However, Gravelle et al. showed that in the presence of reservoirs, excursions of ions outside the pore \LB{coupled to adsorption-desorption processes on the pore surface} result in ionic correlations on much longer time scales, \LB{hence predicting pink noise down to very low frequencies}.

\subsubsection{Wall fluctuations}

So far, we have only considered thermal fluctuations of the fluid itself. However, the confining walls might also be subject to fluctuations. 
Wall fluctuations are known to enhance fluid self-diffusion by inducing flows, through the Taylor-Aris mechanism (\cite{Taylor1953,Aris1956}). On the other hand, static constrictions slow down diffusion in a channel through Fick-Jacobs entropic trapping (\cite{Malgaretti2014,Reguera2001}).
\LB{Recently, MD simulations have shown diffusion enhancement due to the propagation of phonons in carbon nanotubes }(\cite{Ma2015,Cruz-Chu2017}).

On the theoretical side, a very general framework for evaluating the self-diffusion coefficient \NKbis{for a fluid confined in} a fluctuating channel was developed by \cite{Marbach2018}. In a slit like-channel of height $h$, the fluctuation-corrected diffusion coefficient $D$ is formally related \LBbis{to the spectrum of wall fluctuation $S(k,\omega)$} by
\begin{equation}
D = D_0 \left( 1- \frac{1}{h^2} \int \frac{\d k \d \omega}{(2\pi)^2} \frac{(D_0k^2)^2-3 \omega^2}{(D_0k^2)^2+\omega^2} S(k,\omega)\right),
\label{sophie}
\end{equation}
where $D_0$ is the bare diffusion coefficient. Equation (\ref{sophie}) has two limiting regimes, that are governed by a 'fluctuation-related' Peclet number, which may be defined as $Pe = \ell^2/(D_0 \tau)$, where $\ell$ and $\tau$ are the characteristic length and time scales of the fluctuations. The diffusion coefficient is enhanced for $Pe >1$ ($D = D_0(1+3\langle\delta h^2\rangle/h^2)$), and reduced for $Pe <1$ ($D = D_0(1-\langle \delta h^2\rangle/h^2)$) \LBbis{and both tendencies are observed experimentally (\cite{Marbach2018})}. Equation (\ref{sophie}) thus bridges the limiting Taylor-Aris and Fick-Jacobs results. 

\NKbis{The permeability of a fluctuating channel is also expected to be impacted by wall fluctuations. However, permeability is related to the diffusion of the fluid centre of mass (see eq. (\ref{Dcm})), which is different from the fluid self-diffusion; and extending the above framework to centre of mass diffusion remains to be done.}

\subsection{Fluid structuring}
\subsubsection{Phenomenology}

While a fluid appears \LB{disordered} at hydrodynamic lengthscales, its molecular nature manifests itself under confinement in the form of structuring effects. Even at a non-confined solid-liquid interface, the attractive interactions between the solid and the liquid result in molecular layering, as verified experimentally by x-ray spectroscopy (\cite{Cheng2001}). In between two surfaces, a liquid adopts a layered structure even in the absence of interactions, simply due to geometrical constraints (\cite{Israelachvili1983}). The onset of structuring clearly represents the transition from continuum to sub-continuum transport. 

The threshold confinement at which this transition occurs has been studied for both planar and cylindrical geometries using MD simulations. For water in carbon nanotubes, \cite{Thomas2009} have found that it retains a bulk-like disordered structure down to a tube diameter of 1.39 nm ((10,10) tube). In a (9,9) tube (1.25 nm) water was found to structure in stacked hexagons (see figure 7), and in a single file chain in a 0.83 nm (6,6) tube. For water between two graphene sheets, four distinct layers could be observed at a 1.35 nm distance between sheets (\cite{Neek-Amal2016}), and a single monolayer of water was observed for confinement below 0.8 nm (\cite{Neek-Amal2018}). In should further be noted that the observation of an exotic square ice phase for few-layer water between two graphene sheets has been claimed (\cite{Algara-Siller2015}). 

In the sub continuum-regime, transport properties show strong qualitative deviations from bulk expectations. The MD simulations of \cite{Thomas2009} showed non-monotonous permeability for carbon nanotubes as a function of diameter, and similar observations were made for the capillary filling velocity of carbon nanotubes, studied by \cite{Gravelle2016}. 
\LB{On the experimental side, this behaviour can be put in perspective with the recent results by \cite{Radha2016} for capillary-driven flow through slit-like graphene channels }
where a peak in flow rate at around 1.3 nm confinement was observed.

\begin{figure}
\includegraphics[width=\textwidth]{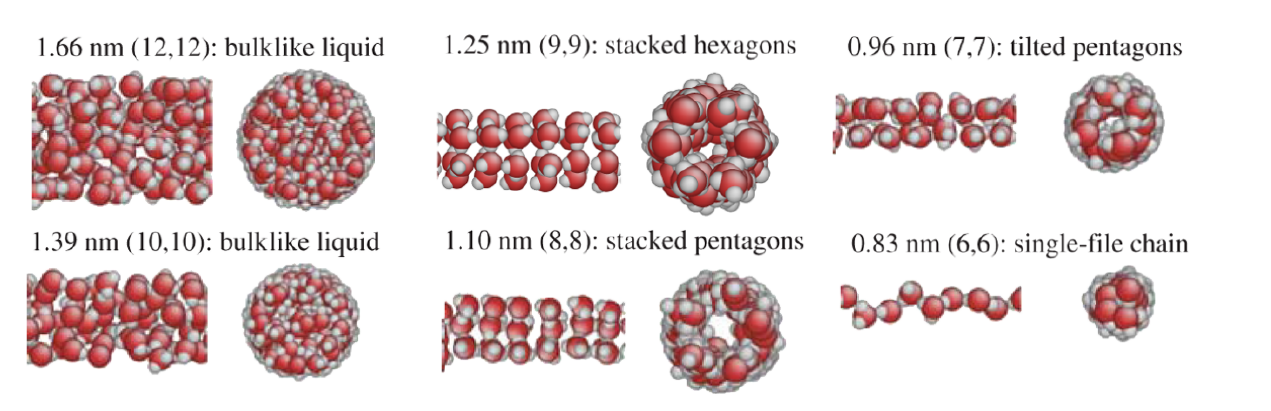}
\caption{MD simulation snapshots of water structure inside carbon nanotubes of different radii, adapted from \cite{Thomas2009}.}
\end{figure}

\subsubsection{Disjoining pressure}
Thermodynamically, the effect of \LB{molecular} fluid structuring may be described by a supplementary contribution to the pressure, the so-called {disjoining pressure}. For a fluid confined between two walls of area $\mathcal{A}$ separated by a distance $h$, in equilibrium with a reservoir at pressure $P_0$, \NKbis{it} is defined in full generality as
\begin{equation}
\Pi_d (h) = - \frac{1}{\mathcal{A}} \left( \frac{\partial G}{\partial h} \right),
\end{equation}
where $G$ is the Gibbs free energy of the whole system (\cite{Israelachvili,Barrat_Hansen}). 

\NKbis{In principle, $G$ varies with $h$ because of distance-dependent interactions between the surfaces; the disjoining pressure can then be seen as the force the fluid needs to exert in order to oppose those interactions. These are generally of two types. On the one hand, there are van der Waals interactions that scale algebraically as $h^{-2}$, yielding a contribution $\Pi_{vdW}=-{A_H\over 6\pi h^{3}}$ to the disjoining pressure, with $A_H$ the Hamaker constant (although more complex dependencies may occur in specific geometries). On the other hand, electrostatic interactions lead to contributions that scale exponentially with the distance $h$ due to ionic screening. The combination of these two interaction terms is the basis for the DLVO (Derjaguin-Landau-Verwey-Overbeek) theory.}

 Now, if there is fluid structuring that depends on confinement, it makes a supplementary contribution to the variation of $G$. This contribution exists even for a hard sphere fluid \LBbis{due to excluded volume effects}, and is purely entropic in nature. It quantifies essentially how much the structure of the fluid in confinement differs from the one in the bulk. 
 For hard spheres of diameter $\sigma$, \LBbis{the disjoining pressure is usually described by the expression }(\cite{Israelachvili,Kralchevsky1995}): 
\begin{equation}
\Pi_d (h) = - \rho_{\infty} k_B T \cos(2\pi h /\sigma) e^{-h/\sigma}, 
\end{equation}
where $\rho_{\infty}$ is the bulk density. This is an oscillating function of $h$ that decays to 0 at large $h$, where there is no more fluid structuring. It is bounded by $\Pi_d(0) = \rho_{\infty} k_B T$, which evaluates to about 1000 bar for water at room temperature. 
In practice, \LBbis{for fluid flow measurements where a pressure drop is imposed between two reservoirs, } this (\NKbis{huge}) contribution to the pressure does not intervene directly. However, in a capillary flow geometry, \LB{which involves liquid-vapour interfaces,} the disjoining pressure may \NKbis{be much larger than the bare capillary pressure and} \LBbis{act as} the main driving force. Indeed, the capillary pressure drop across a meniscus of radius $R = 1~\rm nm$ is typically $2\gamma/R = 140~\rm bar$ (for water with surface tension $\gamma = 72~\rm mN \cdot m^{-1}$). 
\NKbis{Driving by disjoining pressure was observed in simulations of carbon nanotube capillary filling (\cite{Gravelle2016}), where the subtle dependence of the water molecular structure -- and therefore of the disjoining pressure -- on nanotube radius resulted in the imbibition velocity displaying oscillatory behaviour versus the confinement. Experimentally, a similar phenomenology was observed by \cite{Radha2016}, who measured the capillary flow of water across angstrom scale slits and found a non-monotonous dependence of the evaporation rate on confinement. These measurements could be interpreted in terms of disjoining pressure, which was indeed found to be of the order of 1000 bar in MD simulations of the experimental system (\cite{Neek-Amal2018}).}

\subsubsection{Descriptions of fluid transport}
A first understanding of sub-continuum transport may be provided by extending the continuum description with effective, confinement-dependent, values for density, viscosity and slip length. In a layered fluid, the density in each layer is generally higher than in the bulk, and a general trend is that viscosity increases (\cite{Schlaich2017}). More exotically, an oscillating viscosity as a function of confinement was observed in simulations of water between two graphite slabs (\cite{Neek-Amal2016}). 
 \cite{Neek-Amal2018} could reproduce the experimental results of \cite{Radha2016} by using simulated values of disjoining pressure, along with a Poiseuille formula with effective density and viscosity; however, they assumed a viscosity-independent slip length. Now, in general, care should be taken in defining a viscosity for a structured fluid: as its density is non-uniform, a position-dependent viscosity should be introduced in order to describe the details of the flow profile. For confinement below five molecular diameters (in the case of a Lennard-Jones fluid) \cite{Travis1997} have shown that even a position-dependent viscosity is not sufficient and a non-local viscosity kernel should be used (\cite{Zhang2004}). 
 \LB{Ultimately, for confinement below a few molecular layers (typically 1 nm for water), the notion of viscosity itself -- which is intrinsically a continuum quantity --  looses its fundamental meaning.}

Nevertheless, a simple picture of sub-continuum fluid transport is possible in the case of large slippage, \LB{since surface friction then becomes the main} \NK{mechanism resisting fluid transport}.
This \NK{appears indeed when} looking at the limit $b\gg R$ in the Poiseuille formula (\ref{poiseuille}), which yields for the average flow velocity:
\begin{equation}
v = \frac{R b}{2 \eta L} \Delta P = \frac{ R}{2 \lambda L} \Delta P,
\label{finkel}
\end{equation}
where we have obtained the second equality by relating the slip length to the liquid-wall friction coefficient $\lambda$: $b = \eta/\lambda$; \LB{note that, for simplicity, we forget here about entrance.} 
\LBbis{The viscosity does not enter the permeability, as expected for surface dominated friction.}
 The fluid moves indeed in the channel as a single block, \LB{and dissipation only intervenes at the surface}. 
The crucial parameter governing the transport is then the solid-liquid friction coefficient $\lambda$. A model for evaluating $\lambda$, tested against MD simulations of water in carbon nanotubes, was proposed by \cite{Falk2010}. They obtain
\begin{equation}
\lambda \approx \frac{\tau}{k_B T} f_{q_0}^2 S(q_0),
\end{equation}
where $q_0$ is the wave-vector corresponding to the solid lattice spacing, $f_{q_0}$ is the Fourier component of the fluid-solid interaction potential at that wave-vector, and $S(q)$ is the liquid structure factor; $\tau$ is the correlation time of the force between the solid and the liquid, which is found to not depend on confinement. In this way, $\lambda$ is found to be directly related to the liquid structure and to the liquid wall interaction. 
Such a friction-dominated approach was applied successfully, \NKbis{for example}, to predict the permeability of ultra-confined alcanes in a nanoporous matrix, showing a scaling dependence of the permeance on the alcane length (\cite{Falk2015}). 
\LB{However, predicting the solid-liquid friction coefficient remains to large extents an open problem. As we discuss in the following (see 4.2.6), its determination in MD simulations still remains ambiguous quantitatively, and the role of underlying quantum effects should be assessed. }

\subsubsection{Ultimate structuring: single file transport}
The most extreme type of structuring is single-file arrangement of fluid molecules. In the case of water transport, this is the realm of biological channels (\cite{Horner2018a}), where water is conducted through sub-nanometric openings in transmembrane proteins. Single file transport may also occur in artificial channels: it was observed in MD simulations for carbon (\cite{Hummer2001}) and boron nitride (\cite{Won2007}) nanotubes of 0.8 nm in diameter, and experimental observation in carbon nanotubes was recently claimed (\cite{Tunuguntla2017}). Single-file transport is a field of its own and there are dedicated reviews on the subject (\cite{Horner2018a,Kofinger2011}); here we will only sketch the main ideas. 

The notion of single file transport and the underlying exclusion transport models are not restricted to the study of confined fluids. Indeed, a general feature of particles in one dimension that cannot cross each other is sub-diffusive behaviour: the mean squared displacement of a particle scales with time as $\langle \Delta x^2 \rangle \propto t^{1/2}$, as opposed to linear scaling for normal diffusion (\cite{Levitt1973}).
Such anomalous diffusion is encountered in a variety of fields (\cite{Fouad2017}), and may be understood in terms of the normal diffusion of vacancies in the 1D chain; in the case of fluid transport, it was studied by Chou (\cite{Chou1999,Chou1998}). 
\LBbis{However, this effect is not expected to play a decisive role in single-file transport through short channels, where vacancies are unlikely to occur (\cite{Kalra2003}).}

\NKbis{In the case where no vacancies are expected, a simple model} \NK{for single-file transport was proposed decades ago by \cite{Finkelstein1974} for describing water transport in biological channels.} It stems from a global force balance on the water chain, and is as such equivalent to eq. (\ref{finkel}), which we have derived as a strongly confined limit of the Poiseuille formula. In Finkelstein's model, the chain of $N$ water molecules (of total length $L$) moving at velocity $v$ is subject to a friction force $- N \xi v$, and to a pressure driving $N v_w \Delta P /L$, where $v_w$ is the volume of one water molecule. The force balance then leads to 
\begin{equation}
v = \frac{v_w}{\xi L} \Delta P,
\end{equation}
which is indeed eq. (\ref{finkel}), granted the identification $\xi = (2 v_w/R) \lambda$. 
In the biophysics literature, the preferred quantity for characterising a channel's transport properties is the unitary channel permeability $p_f$: $p_f$ is the number of water molecules that crosses the channel per unit time, per unit osmolyte concentration difference applied across the channel, therefore expressed in $\rm m^3\cdot s^{-1}$. For an osmolyte concentration difference $\Delta \rho_s$, the osmotic pressure difference is $k_B T \Delta \rho_s$, and hence the unitary permeability is 
\begin{equation}
p_f = \frac{N}{L} \frac{v}{\Delta \rho_s} = \frac{v_w}{a L} \frac{k_B T}{\xi},
\end{equation}
where we have defined $a = L/N$ the average spacing between water molecules. The diffusion coefficient of the water chain centre of mass may be expressed through the Einstein relation as $D = k_B T/(N\xi)$, and therefore $p_f = v_w D/a^2$, hence the diffusion coefficient and the unitary permeability may be used interchangeably. \LB{A typical value for the permeability in biological channels such, as Aquaporin or Gramicidin A, is in the range of $10^{-14}-10^{-13}~\rm cm^3 \cdot s^{-1}$}, \NK{which corresponds to 1-10 water molecules crossing the channel per nanosecond under a 1 bar pressure drop.}

\NK{In an alternative phenomenological description, transport through the channel is viewed as an activated process with activation energy $\Delta G^{\ddagger}$ (\cite{Horner2018a}). The permeability is then expressed in the framework of transition state theory as $p_f=v_w\,\nu_0\, \exp[-\Delta G^\ddagger/k_BT]$, with $\nu_0 \sim 10^{13}$ s$^{-1}$ a molecular attempt frequency. }
For biological channels, this relation is well verified by independent measurements of $p_f$ and $\Delta G^\ddagger$, with \NKbis{$\Delta G^\ddagger$ of the order of $ 5~\rm kcal \cdot mol^{-1}$}. \LBbis{However, values of $\Delta G^\ddagger$ remains debated for transport measurements in carbon nanotubes (\cite{Horner2018,Tunuguntla2018}). }

Although it provides a general guiding line, Finkelstein's formula is challenged both in artificial and biological channels. In the case of protein channels, the assumption of uniformly smooth walls breaks down, since there are discrete hydrogen bonding sites (\cite{Horner2018a}). Hence $p_f$ is not found to be inversely proportional to channel length (\cite{Saparov2006}): it rather has an exponential dependence on the number of hydrogen bonding sites (\cite{Horner2015}). Such an exponential dependence suggests a collective transport mechanism of the water chain, with bursts requiring the breaking of multiple hydrogen bonds at once. There is, however, a notable disagreement between experiments and simulations, as the simulations of water transport through polyalanine channels showed no dependence of $p_f$ on the channel length (\cite{Portella2007}). 

In simulations of carbon nanotube channels, all the single-file water molecules were clearly shown to move in a correlated fashion (\cite{Hummer2001}): water transport occurs when all the molecules simultaneously shift by one molecular diameter. These dynamics were successfully described by a continuous time random walk model (\cite{Berezhkovskii2002}), or equivalently by diffusion of a collective coordinate of the water molecules (\cite{Zhu2004}), which actually echoes our discussion of fluctuations in a more general setting (eq. \ref{langevin} and \cite{Detcheverry2012}). 

Overall, there is still much to understand about the collective motions and subtle surface interactions involved in single-file transport, and it is an active field of research. 

\subsubsection{Structuring and electrostatics: dielectric anomalies}

\begin{figure}
\includegraphics[width=\textwidth]{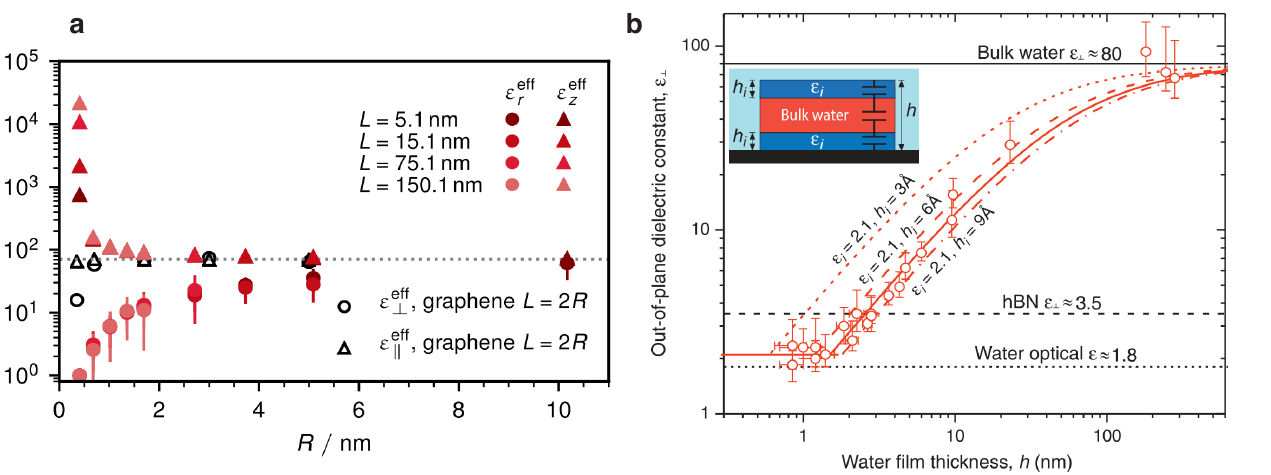}
\caption{Dielectric anomalies due to fluid structuring. \textbf{a.} Components of the dielectric permittivity tensor of water in a carbon nanotube, as a function of nanotube radius, as determined from MD simulations by \cite{Loche2019}. \textbf{b.} Experimental results for the transverse dielectric constant of water in planar confinement, as a function of confinement width, from \cite{Fumagalli2018}.}
\end{figure}

As confined water becomes structured, its response properties to an external electric field are accordingly modified. When an electric field $\bf E$ is applied in water, the individual molecules reorient and polarise, creating an electric field themselves. The total electric field $\bf E$ is then the sum of the polarization field of the water molecules and of the externally applied field $\mathbf{D}/\epsilon_0$, where $\bf D$ is called the electric displacement and $\epsilon_0$ is the vacuum permittivity. In bulk water, the dielectric response, that is the relation between $\bf D$ and $\bf E$, may be expressed through a single scalar quantity, the relative permittivity $\epsilon \approx 80$: $\mathbf{D} = \epsilon \epsilon_0 \mathbf{E}$. However, the most general (static) linear response may be anisotropic, space-dependent and non-local: 
\begin{equation}
D_{\alpha}(\mathbf{r}) = \epsilon_0 \sum_{\beta} \int \d \mathbf{r}' \epsilon_{\alpha \beta} (\mathbf{r},\mathbf{r'}) E_{\beta} (\mathbf{r'}).
\end{equation}
The relative permittivity is then a tensor with components $\epsilon_{\alpha \beta} (\mathbf{r}, \mathbf{r}')$. While MD simulations show that the dielectric response in water may be considered local ($\epsilon_{\alpha \beta} (\mathbf{r}, \mathbf{r}') = \epsilon_{\alpha \beta} (\mathbf{r}) \delta(\mathbf{r}-\mathbf{r'})$), it becomes anisotropic and space-dependent in the vicinity of interfaces, as a consequence of the water layering (\cite{Bonthuis2012},\cite{Bonthuis2011}). Qualitatively, the orientations of the water dipoles are anti-correlated in the direction perpendicular to the interface, resulting in a reduced permittivity in that direction, while the permittivity is largely unaffected parallel to the interface. In planar confinement, this behaviour could be captured by an effective medium model, in which the water is described by a space-independent, but anisotropic permittivity $(\epsilon_{\parallel},\epsilon_{\perp})$ (the parallel direction is not confined). While $\epsilon_{\parallel}$ essentially retains its bulk value, $\epsilon_{\perp}$ is reduced by up to an order of magnitude for confinements below 1 nm (\cite{Schlaich2016,Zhang2013}). Such a reduction of the perpendicular dielectric response was recently observed experimentally for water confined between a graphite and a boron nitride crystal (\cite{Fumagalli2018}). A deviation from the bulk value was measured up to nearly 100 nm confinement (figure 7). The results were well described by assuming each interface carried a 7 \AA ~thick layer of very low permittivity ($\epsilon = 2.1$) "electrically dead water". 

An effective medium model based on MD simulations was developed by \cite{Loche2019} for cylindrical confinement of water in carbon nanotubes. Similarly to the case planar case, the radial permittivity is found to be reduced by up to an order of magnitude for tube radii smaller than 3 nm (figure 7). However, for the smallest tubes (below 1 nm radius), the longitudinal permittivity is found to increase with respect to its bulk value, and it skyrockets 1 to 3 orders of magnitude in a 0.4 nm radius tube, where water is in a single file arrangement. 

Knowledge of these modified dielectric properties is important when considering ion transport in strong confinement, as will be discussed in the following (see 4.3.1). 

\subsubsection{Limits of molecular dynamics}

We will close this section on liquid structuring with a word of warning. Many of the results we have presented so far have been obtained using classical MD simulations, and like any type of simulation, these come with some underlying assumptions. In particular, as we are dealing with strongly confined fluids where all the molecules interact with the confining surface, the modelling of that surface plays an increasingly important role. Since the solid is represented by rigid balls interacting with classical force fields, no electron dynamics may be described by the simulation. However, some coupling between water and electron dynamics may be expected, based on a lengthscale argument. The maximum of the dielectric response of a solid indeed occurs at a lengthscale $k_{\rm TF}^{-1}$, where $k_{\rm TF}$ is the Thomas-Fermi wave-vector (\cite{Mahan}). Typically, $k_{TF}^{-1} \approx 1~\rm nm$ in graphite, hence the electrons in graphite could expected to respond to the dynamics of individual water molecules. 

There is growing evidence that such electron-water couplings indeed occur. A first consequence of taking into account electron dynamics is that a solid should appear polarisable: in simulations, the fluctuations of water next to a graphite surface were found to be strongly affected by the polarisability of carbon atoms (\cite{Misra2017}). On the experimental side, the induction of an electronic current by a water flow inside a carbon nanotube, as well as the reverse phenomenon, were observed (\cite{Ghosh2003,Rabinowitz2020}). 

These results call for caution in the interpretation of MD simulations, which, as we have pointed out, lack electron dynamics. They also call for the development of new simulation methods that would take such dynamics into account, while not limited by the very small system size of \emph{ab initio} simulations. 

\subsection{Electrostatics in extreme confinement}
In this section, we discuss a few aspects of the behaviour of ions below the continuum limit. We will describe how the confinement modifies ionic interactions, and possible consequences on ion transport. 

\subsubsection{Ionic interactions and self-energy}	

In bulk water, ions interact via a Coulomb potential $\phi(r) = e/(4\pi \epsilon_0 \epsilon_w r)$, where $\epsilon_w$ is the water dielectric permittivity. But inside a nanochannel, ions are no longer surrounded by a homogeneous fluid, and their interaction potential may be affected by the dielectric properties of the confining medium. In order to assess the importance of this effect, we consider the simple situation where an infinitely long cylindrical channel of radius $R$ is filled with water having isotropic permittivity $\epsilon_w$, and the medium outside the channel is a homogeneous dielectric of permittivity $\epsilon_s$, with typically $\epsilon_s \ll \epsilon_w$. The electrostatic potential around an ion placed in the middle of the channel can then be determined by solving Poisson's equation in the presence of the dielectric discontinuities, which has been done analytically by several authors (\cite{Parsegian1969,Levin2006,Teber2005,Kavokine2019}). 

The complicated analytical result may be interpreted in the following way (see figure 8). At short distances (much smaller than the channel radius), only the dielectric response of the water is visible and the potential is $\phi(x) \sim 1/(\epsilon_w x)$. At long distances (much larger than the channel radius), it is the dielectric response of the confining medium that matters and $\phi(x) \sim 1/(\epsilon_s x)$. At intermediate distances, the electric field lines are essentially parallel to the channel due to the dielectric contrast $\epsilon_w \gg \epsilon_s$, and the potential, which is well described by an exponential function, resembles a 1D Coulomb potential: 
\begin{equation}
\phi(x) \approx \frac{e \alpha}{2 \pi \epsilon_0 \epsilon_w R} e^{-x/(\alpha R)},
\label{1DC}
\end{equation}
with $\alpha$ a numerical coefficient that depends on the ratio $\epsilon_w/\epsilon_s$; $\alpha = 6.3$ for $\epsilon_w/\epsilon_s = 40$ (\cite{Teber2005}). It is notable that so far no assumption on the channel radius was made, so that formally this 1D regime exists for a channel of any size. However, it is only relevant if it leads to ion-ion interactions stronger than $k_BT$, that is if $e \phi(R) > k_B T$, for a monovalent ion. This defines a limiting channel radius, below which ionic interactions are affected by the confining medium: $R_c \sim 7.5~\rm nm$. 
These modified Coulomb interactions therefore may have some effect in any {single-digit nanopore}, and they have essentially two practical consequences. 

\begin{figure}
\includegraphics[width=\textwidth]{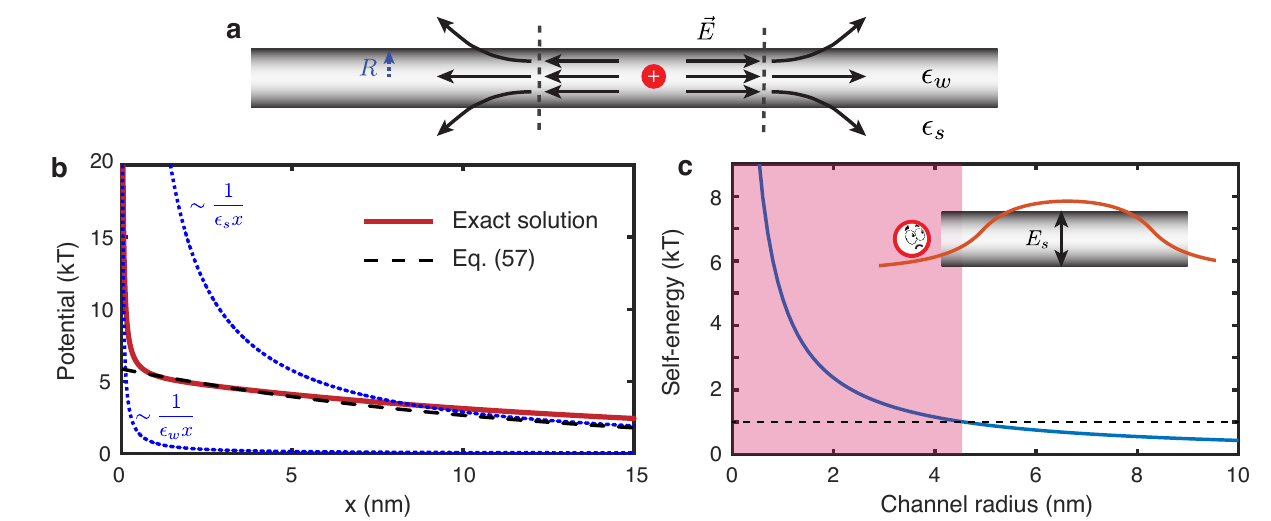}
\caption{\textbf{a.} Schematic representation of the electric field lines around an ion confined in a narrow channel. \textbf{b.} Potential generated along the axis by an ion confined in a channel of radius 2 nm. \textbf{c.} Born self-energy of a monovalent ion as a function of channel radius, computed following \cite{Teber2005}.}
\end{figure}

First, because ions produce a stronger Coulomb potential in confinement than in the bulk, they have a larger Born self-energy. This phenomenon was pointed out decades ago by \cite{Parsegian1969} in the case of ions crossing lipid bilayers, and has since then been commonly considered in the study of biological ion channels. For small channels where the 1D contribution to the potential dominates, the additional self-energy due to confinement is ${\cal E}_s = e\phi(0)/2$, with $\phi(x)$ given by eq. (\ref{1DC}). An expression valid for any $R$ was obtained by Teber, and it is plotted in figure 8. It shows that ${\cal E}_s$ is larger than $k_B T$ up to $R \sim 4~\rm nm$. ${\cal E}_s$ represents a supplementary energy barrier for entering the channel. \LBbis{To illustrate the consequence of this Born self-energy, let us consider} a neutral channel in the absence of correlation effects: a barrier ${\cal E}_s = 4k_BT$ reduces the ionic concentration by a factor $e^{{\cal E}_s/(k_BT)} = 100$. 

Second, increased Coulomb interactions with respect to the bulk may actually result in correlation effects. To assess the potential importance of correlations, one may introduce a coupling parameter $\Gamma = e \phi((\pi R^2 \rho)^{-1})/(k_B T)$, where $\rho$ is the ion concentration inside the channel, so that $(\pi R^2 \rho)^{-1}$ is the average distance between ions (\cite{Bocquet2010}). Let us take as an example a channel of radius 2 nm, where ${\cal E}_s \sim 2.5 k_BT$ from eq.~(\ref{1DC}). An effective Coulomb potential of magnitude $\sim 5 k_B T$ then extends around an ion over a distance $\alpha R \sim 10~\rm nm$. Therefore, for concentrations above $\rho = 10^{-2}~\rm M$, $\Gamma > 1$ and correlations are expected to become important. 

Several remarks can be made here. A first point is that we have considered monovalent ions so far; for ions of valence $z$, the self-energy, for example, would be multiplied by a factor $z^2$. Then, we have gone through the discussion assuming for simplicity that water has an isotropic permittivity $\epsilon_w$. We have highlighted in the previous section that this assumption breaks down in cylindrical channels of radius smaller than 5 nm, and, though it does not affect the qualitative phenomenology, dielectric anisotropy, as predicted by \cite{Loche2019}, should then be taken into account. It mainly affects the behaviour of the potential at short distances $x \leq R$. 
We should also mention that when entering a channel smaller than its hydrated radius, an ion pays an additional energy penalty due to the shedding of its hydration shell (\cite{Richards2012,Epsztein2019}). For chloride, the total hydration energy is as high as $155~k_B T$, but the hydrated radius is 0.4 nm, so that only partial dehydration arises and only for the smallest pores. 
Finally, we have assumed that the confining medium is a dielectric medium with uniform permittivity. This is a good model for a lipid membrane, but artificial confining materials may have a variety of electronic properties (see 4.2.6), and their influence on effective Coulomb interactions has not been thoroughly investigated. In the case of confinement by a perfect metal, the ionic interactions were found to be exponentially screened over a distance of order of the channel radius (\cite{Kondrat2011,Loche2019}). Such effective interactions were found to lead to like-charge attraction in ionic liquids confined in carbon nanopores (\cite{Futamura2017}). 

\subsubsection{Ion transport beyond the mean-field}

We have highlighted in the previous paragraph that correlations may become important in ion transport through single digit nanopores. If it is the case, then mean-field theories, such as the PNP framework, may not be directly applied, and the determination of equilibrium properties, such as ion concentrations inside the nanopore, requires some form of exact statistical mechanics. In systems with high aspect ratio (typically nanotubes), the problem can be reduced to a one-dimensional model of the Ising or 1D Coulomb gas type, and the partition function may then be exactly computed through a (functional) transfer matrix formalism (\cite{Zhang2006,Lee2014b,Kavokine2019}). Outside of the 1D geometry, variational methods may be used (\cite{Buyukdagli2010}), and methods for incorporating ion pairs into Poisson-Boltzmann theory have been developed (\cite{Y.Levin2002}). A general feature of these calculations (\cite{Zhang2006,Buyukdagli2010}) is that they predict filling transitions: namely, a non-analytic behaviour of ion concentrations in the nanopore as a function of the salt concentration in the reservoir, which strongly deviates from the mean-field Donnan equilibrium result, and may break local electroneutrality (\cite{Zhang2006}). One should note, however, that breakdown of electroneutrality does not require ionic correlations and it has recently been shown to arise in a mean-field setting (\cite{Levy2020}). 

If linear response theory applies, then transport properties such as ionic conductance may be determined directly from the concentrations of charge carriers at equilibrium. However, this is no longer the case when correlations are very strong, leading to the formation of tightly bound Bjerrum pairs of oppositely charged ions. This was first realised by Onsager, who showed that in a three-dimensional weak electrolyte -- an electrolyte that forms Bjerrum pairs -- there is a quadratic current-voltage relation, a phenomenon known as the second Wien effect (\cite{Onsager1934,Kaiser2013}). Bjerrum pairing in one-dimensional confinement was studied by \cite{Kavokine2019}, and stable ion pairs were shown to arise typically for confinement below 2 nm, similarly to what was observed in MD simulations (\cite{Nicholson2003}). The pairing resulted in some very non-linear behaviour governed by discrete particle effects, \LB{echoing} the phenomenology of Coulomb blockade (\cite{Beenakker1991,Kaufman2015}). \NK{As such, ion correlations produced quantised transport behaviour in a purely classical system.} 

The coupling of ion transport to fluid transport in the non-mean-field, non linear regime remains largely unexplored, though it promises a rich phenomenology. As an example, a recent simulation (\cite{Li2017}) found that pressure driven water flow through a nanochannel could be blocked by an ion tightly bound inside. Lastly, we should mention, at the frontier of sub-continuum ion transport, the Grotthus-like translocation of protons, which has been observed in a single file (\cite{Tunuguntla2017}) and a single plane (\cite{Gopinadhan2019}) of water, but still remains poorly understood \LBbis{in strongly confined systems}.

\section{CONCLUSION}

This review has explored some defining aspects of fluid transport at the nanoscale. We started from continuum theory and reduced the scale down to its limit of applicability; we then explored phenomena that occur below the continuum limit and gave indications of appropriate theoretical descriptions to tackle them. The main points summarised below were made. Overall, fluids in molecular scale confinement are largely an uncharted territory for theory, and recent experiments urge for the development of theoretical tools beyond those described in this review. 
\begin{summary}[SUMMARY POINTS]
\begin{enumerate}
\item Experimental systems for studying fluid transport in molecular scale confinement are today within reach. 
\item Above 10 nm confinement, fluid transport is governed by continuum hydrodynamic equations, with coupling to ion transport and surface effects. 
\item Below 10 nm --the domain of so-called single-digit nanopores -- thermal fluctuations and electrostatic correlations are increasingly important, challenging continuum and mean-field theory. 
\item In few nanometre confinement, fluid structuring effects and correlations play an overwhelming role. 
\end{enumerate}
\end{summary}

\section*{DISCLOSURE STATEMENT}
The authors are not aware of any affiliations, memberships, funding, or financial holdings that
might be perceived as affecting the objectivity of this review. 

\section*{ACKNOWLEDGMENTS}
L.B. acknowledges funding from the EU H2020 Framework Programme/ERC
Advanced Grant agreement number 785911-Shadoks. L.B. and R.N acknoweldge support from ANR-DFG project
Neptune. R.N. acknowledges funding from the DFG via SFB1078 and NE810/11. N.K. thanks A. Marcotte for discussions.
%

\newpage

\section*{{\Large Appendix}}

\begin{table}[h]
\begin{tabular}{|c|c|}
\hline
$\rho_s$ (M) & $\lambda_D$ (nm)\\
\hline
$10^0$ & 0.3 \\
$10^{-1}$ & 1.0 \\
$10^{-2}$ & 3.1 \\
$10^{-3}$ & 9.6 \\
\hline
\end{tabular}
\begin{tabular}{|c|c|c|c|}
\hline
Material &  $\Sigma$ max ($ e\rm\cdot nm^{-2}$)  & $\ell_{GC}$ (nm) & $\ell_{Du}$ at 0.1 M (nm)\\
\hline
 Graphene oxide$^{\rm a}$  & 0.06 & 4 & 9  \\
Silica$^{\rm b}$ & 0.3 & 0.8 & 50 \\
Carbon nanotube$^{\rm c}$ & 1 & 0.2 & 170 \\
BN nanotube$^{\rm d}$ & 9 & 0.03 & 1500 \\
\hline
\end{tabular}
\begin{tabnote}
$^{\rm a}$\cite{Mouhat2020}; $^{\rm b}$\cite{Stein2004};$^{\rm c}$\cite{Secchi2016a}; $^{\rm d}$\cite{Siria2013}
\end{tabnote} 
\caption{Electrostatic lengthscales}
Left: values of Debye length for different concentrations of monovalent salt. Right: maximal measured surface charge, Gouy-Chapman length and Dukhin length at 1 M monovalent salt concentration.
\end{table}

\section*{Surface PNP theory}

Here we reproduce the main equations of the surface PNP-Stokes framework, that was introduced by \cite{Mouterde2019} (see 3.3.5). 
The Nernst-Planck expression for the ion fluxes is modified according to 
\begin{equation}
j_{\pm} = \frac{e}{\xi_{\pm}+\lambda_{\pm}} \left[ \frac{k_B T}{e} (-\partial_x \rho_{\pm} \pm \rho_{\pm} (-\partial_x \phi) \right] + \alpha_{\pm}\rho_{\pm} v,
\end{equation}
with $\alpha_{\pm} = \frac{\xi_{\pm}}{\xi_{\pm}+\lambda_{\pm}}$. The (integrated) Stokes equation becomes
\begin{equation}
v = K(\rho_+,\rho_-) [ (-\partial_x p)+ e(\alpha_+ \rho_+- \alpha_-\rho_-)(-\partial_x \phi)],
\end{equation}
with $K(\rho_+,\rho_-)$ a concentration-dependent permeability, and these are supplemented by a 1D Poisson equation: 
\begin{equation}
\partial_x[\epsilon h (-\partial_x \phi)] = h e (\rho_+-\rho_-).
\end{equation}

\end{document}